\documentclass[aps,prx,onecolumn,superscriptaddress,longbibliography]{revtex4-2}

\usepackage{amsmath,amsfonts,amsthm,bbm}
\usepackage[dvipsnames]{xcolor}
\usepackage{graphicx}
\usepackage{multirow,chemformula}
\usepackage[version=4]{mhchem}
\usepackage[colorlinks=true,breaklinks=true,linkcolor=magenta,citecolor=magenta,urlcolor=magenta]{hyperref}

\usepackage{tikz}
\usetikzlibrary{quantikz}

\newcommand{\crt}[1]{\hat{c}_{#1}^\dagger}
\newcommand{\dst}[1]{\hat{c}_{#1}^{\phantom{\dagger}}}
\newcommand{\vett}[1]{{\bf{#1}}}
\newcommand{\bgreek}[1]{{\boldsymbol{#1}}}
\newcommand{\ry}{R_y}
\newcommand{\ang}{\mathrm{\AA}}

\begin{document}

\title{Challenges in the use of quantum computing hardware-efficient Ans\"{a}tze \\ in electronic structure theory}

\author{Ruhee D'Cunha}
\affiliation{Department of Chemistry, Virginia Tech, Blacksburg, VA 24061, USA}

\author{T. Daniel Crawford}
\affiliation{Department of Chemistry, Virginia Tech, Blacksburg, VA 24061, USA}

\author{Mario Motta}
\affiliation{IBM Quantum, IBM Research Almaden, 650 Harry Road, San Jose, CA 95120, USA}

\author{Julia E. Rice}
\email{jrice@us.ibm.com}
\affiliation{IBM Quantum, IBM Research Almaden, 650 Harry Road, San Jose, CA 95120, USA}

\begin{abstract}
Advances in quantum computation for electronic structure, 
and particularly heuristic quantum algorithms, 
create an ongoing need 
{to characterize the performance and limitations of these methods.
Here we discuss some potential pitfalls connected with the use of hardware-efficient 
Ans\"{a}tze in variational quantum simulations of electronic structure.
We illustrate that hardware-efficient Ans\"{a}tze may break Hamiltonian symmetries and
yield non-differentiable potential energy curves, in addition to the well-known difficulty of optimizing variational parameters.
We discuss the interplay between these limitations by carrying out a comparative analysis of hardware-efficient Ans\"{a}tze versus unitary coupled cluster and full configuration interaction, and of second- and first-quantization strategies to encode fermionic degrees of freedom to qubits.
Our analysis should be useful in understanding potential limitations and in identifying possible areas of improvement in hardware-efficient Ans\"{a}tze.}
\end{abstract}

\maketitle

\section{Introduction}

Recent years have witnessed remarkable research in the simulation of many-body quantum systems by quantum computing algorithms \cite{georgescu2014quantum,cao2019quantum,bauer2020quantum,mcardle2020quantum,motta2021emerging,cerezo2020variational}.
A prominent application targeted by quantum algorithms is the electronic structure (ES) problem, 
namely solving for the ground- or low-lying eigenstates of the electronic Schr\"{o}dinger equation for atoms, molecules, and materials.
Research in this direction has delivered important results in the calculation of potential energy curves, ground- and excited-state energies, and correlation functions \cite{o2016scalable,kandala2017hardware,google2020hartree,rice2021quantum} for various electronic systems.

{
Variational quantum simulations are among the most widespread strategies to carry out ES calculations on quantum computers today \cite{peruzzo2014variational,cerezo2020variational}. These include the variational quantum eigensolver (VQE) algorithm \cite{peruzzo2014variational}. Within VQE, one defines an Ansatz approximating the ground state of a target Hamiltonian, i.e. a set of wavefunctions obtained by applying a parametrized quantum circuit to an input state. 
Central to the accuracy and computational cost of a variational quantum simulation is the construction of an Ansatz with which to explore the space of approximate solutions to the target problem.

Hardware-efficient Ans\"{a}tze \cite{kandala2017hardware} are families of wavefunctions designed with the primary goal of being compatible with the budget of near-term quantum hardware in terms of qubit number and connectivity, native gates of the device, and circuit depth.
While hardware-efficient Ans\"{a}tze are a useful support in the demonstration of VQE and other variational quantum algorithms, their application to the ES problem remains to be fully understood and established.

In this study, we explore some potential pitfalls in the use of quantum computing hardware-efficient Ans\"{a}tze in ES theory, using deviations between the computed and exact (or full Configuration Interaction, FCI) properties as an indicator of algorithmic quality.
We observe that hardware-efficient Ans\"{a}tze 
%may
can
break Hamiltonian symmetries, a conceptually important but often overlooked aspect of quantum simulations. Furthermore, shallow Ans\"{a}tze can yield 
non-differentiable potential energy curves, leading to ill-defined nuclear forces.
Furthermore, the optimization of variational parameters is a challenging and expensive operation. Indeed, it is well-known that hardware-efficient Ans\"{a}tze can exhibit barren plateaus, where under certain conditions the gradient of the cost function vanishes exponentially with system size \cite{mcclean2018barren}. 
}

The remainder of this work is organized as follows: after describing the methods used in this work, we present and discuss the results of this study. Supporting Information (SI) contains additional computational and algorithmic details, together with the schema of a repository which can be used to store results from this study and other studies that evaluate the properties of new hardware Ans\"{a}tze.

\section{Methods}
\label{sec:methods}

In this section, we describe the systems and methods studied in this work, the potential pitfalls of hardware-efficient Ans\"{a}tze in the study of ES, and the metrics used to assess them.

\subsection{Target molecules}

In this work, we focused on a set of five molecular species, namely LiH, BH, HF, BeH$_2$, and H$_2$O.
The LiH,  BH, HF, BeH and OH bond lengths $R$ varied between 0.9-5.3, 0.7-3.5, 0.7-4.5, 0.7-4.5, and 0.7-3.3 $\ang$ respectively.
These molecular species, proposed in previous work \cite{kandala2017hardware,gomes2020efficient,rice2021quantum}, 
allow testing of hardware-efficient variational Ans\"{a}tze along the breaking of one (LiH, BH, and HF) or 
two (BeH$_2$, and H$_2$O) covalent bonds, and in the presence of dominant dynamic 
(small $R$) or static (medium to large $R$) electronic correlation.

For each species, a restricted closed-shell Hartree-Fock calculation was performed, 
as specified in Section II of the SI. The Born-Oppenheimer Hamiltonian
\begin{equation}
\label{eq:hamiltonian}
\hat{H} = E_0 + \sum_{\substack{p r \\ \sigma}} h_{p q} \crt{p \sigma} \dst{q \sigma} + \sum_{\substack{p q r s \\ \sigma\tau}} \frac{(pr|qs)}{2} \crt{p \sigma} \crt{q\tau} \dst{s\tau}  \dst{r\sigma}
\end{equation}
was computed, along with auxiliary operators $\hat{X}$, i.e. the number (of electrons), spin-$z$, and total spin operators
\begin{equation}
\label{eq:auxiliary}
\hat{N} = N_f + \sum_{p \sigma} \crt{p \sigma} \dst{p \sigma} \quad,\quad 
\hat{S}_z = \sum_{p \sigma} (-1)^\sigma \crt{p \sigma} \dst{p \sigma} \quad,\quad
\hat{S}^2 = \left[ \sum_p \crt{p\downarrow} \dst{p\uparrow} \right] \left[ \sum_q \crt{q\uparrow} \dst{q\downarrow} \right]  + \hat{S}_z (\hat{S}_z+1) \quad,
\end{equation}
where $N_f$ is the number of electrons in frozen (closed-shell) orbitals.
{In this study, we used a minimal STO-6G basis. While these problems are notoriously modest in size, they allow to observe and discuss important potential limitations of hardware-efficient variational Ans\"{a}tze.}

\subsection{Qubit encodings}

In the majority of quantum algorithms for contemporary devices, the Hamiltonian Eq.~\eqref{eq:hamiltonian} and the auxiliary operators Eq.~\eqref{eq:auxiliary} 
are mapped onto linear combinations 
\begin{equation}
\label{eq:qubit}
\hat{H} = \sum_{i=0}^{n_p-1} c_i \, \sigma_{\vett{v}_i \vett{w}_i} 
\quad,\quad
\sigma_{\vett{v}\vett{w}} = \bigotimes_{k=0}^{n_q-1} \sigma_{v_kw_k} 
\quad,
\end{equation}
of Pauli operators acting on a certain number $n_q$ of qubits. 
In Equation~\eqref{eq:qubit} $\sigma_{00,01,10,11} = \mathbbm{1},X,Z,Y$ is a 
single-qubit Pauli operator, $c_i$ a real-valued coefficient, and the summation 
ranges over $n_p$ terms.
The rationale behind this choice is that a Pauli operator acting on $k$
qubits can be exponentiated with a circuit of up to $4k-1$ gates and measured 
using a single layer of $k$ single-qubit gates, which makes the representation
Eq.~\eqref{eq:qubit} suitable for near-term quantum devices.

In this study, we employed both first-quantization and second-quantization encodings.
First-quantization establishes a one-to-one mapping between the Hilbert space spanned by a set of $K$ space- and spin-adapted configuration state functions (CSFs) and the Hilbert space of $n_q = \lceil \log_2 K \rceil$ qubits.
Second-quantization establishes a one-to-one mapping between the Fock space of electrons in $M$ spatial orbitals and the Hilbert space of $n_q = 2M$ qubits.
{First-quantization encodings automatically ensure access to symmetry-adapted many-electron wavefunctions} and require fewer qubits than their second-quantization counterparts since  $\lceil \log_2 K \rceil \leq 2 M$.
Furthermore, first-quantization encodings lead to sparse operators which can be efficiently exponentiated \cite{aharonov2003adiabatic}, and the exponential operators can then be used to measure their expectation values with the quantum phase estimation algorithm \cite{kitaev1995quantum}.
Notwithstanding these desirable features of first-quantization encodings, the exponentiation of sparse operators and the use of such exponentials inside quantum phase estimation or variational algorithms are expensive operations, beyond the capabilities of near-term quantum devices.
On these devices, operators are more conveniently expressed as linear combinations of a certain number $n_p$ of Pauli operators acting on $n_q$ qubits. However, such first-quantization
representations lead to an $n_p$ that scales exponentially with $n_q$. {This translates to a polynomial scaling with $K$, which in turn increases combinatorially with system size.}
Additional technical details are provided in the following Subsection.
%Section \ref{sec:first}.

Second-quantization encodings give rise to a compact representation of the Hamiltonian and of auxiliary operators Eq.~\eqref{eq:auxiliary}
as linear combinations of a polynomial number of Pauli operators \cite{seeley2012bravyi,bravyi2002fermionic}, but allow the possibility that Hamiltonian 
symmetries are broken (vide infra).
%, for multiple reasons discussed in Section \ref{sec:metrics}.
The issue of symmetry-breaking can be alleviated in various ways. For example, one can
restrain the expectation value of a target observable (e.g. spin) to a desired value
by performing a variational optimization that includes a penalty term. While this strategy
does not ensure access to an eigenstate of the target observable, deviations from an
eigenstate can be assessed by computing variances, for example.
Another way to alleviate symmetry-breaking is to project a wavefunction inside a target
subspace of the Hilbert space. This projection removes symmetry-breaking by construction and allows assessment of the severity of symmetry-breaking by computing the expectation 
value of the projector.

In this work, in second-quantization simulations, we alleviated symmetry-breaking by combining 
the parity mapping with the two-qubit and tapering techniques \cite{bravyi2017tapering,setia2019reducing}. 
The former removes two qubits to ensure the conservation of the parity operators
$(-1)^{\hat{N}_\uparrow}$ and $(-1)^{\hat{N}_\downarrow}$, and the latter identifies a subgroup 
of the molecular orbital symmetry group isomorphic to $\mathbb{Z}_2^{\times k}$ and removes $k$
qubits to ensure the simulated wavefunctions lie in a target irreducible representation 
of the identified abelian subgroup \cite{bravyi2017tapering,setia2019reducing}.
The species studied here have molecular orbital symmetry groups 
($C_{\infty v}$ for LiH, BH and HF, $D_{\infty h}$ for BeH$_2$ and $C_{2v}$ for H$_2$O) isomorphic to $\mathbb{Z}_2 \times \mathbb{Z}_2$ so that the combined use of two-qubit reduction and tapering leads to $n_q = 2M-2$ for LiH (since the calculations on LiH used only $4$ spin orbitals arising from $2s$ and $2p_z$), $n_q = 2M-5$ for BeH$_2$, and $n_q = 2M-4$ otherwise.

\subsubsection{First-quantization encoding}
\label{sec:first}

To carry out simulations in first-quantization, we employ  \cite{fischer2019symmetry,sugisaki2018quantum,sugisaki2019open,gunlycke2020efficient,gunlycke2021compact}
a basis of Slater determinants $\{ {\bf{x}}_i \}_{i=0}^{n_c-1}$ of $n_e = N_\alpha+N_\beta$ electrons in $M$ spatial orbitals
(specifically the MOs from a restricted closed-shell HF simulation). 
The number of such determinants is $n_c = \binom{M}{N_\alpha} \binom{M}{N_\beta}$.
We then select determinants in the same irreducible representation (irrep) of the molecular orbital symmetry group to which the HF state belongs, i.e. the totally symmetric irrep in the Abelian point group relevant to the present work.
In the basis of such determinants, we construct the Hamiltonian and total spin matrices
\begin{equation}
H_{ij} = \langle {\bf{x}}_i | \hat{H} | {\bf{x}}_j \rangle
\quad,\quad
S_{ij} = \langle {\bf{x}}_i | \hat{S}^2 | {\bf{x}}_j \rangle
\quad.
\end{equation}
We diagonalize the total spin matrix obtaining a basis of configuration state functions (CSFs), $S_{ij} \phi_{j\mu} = s_\mu \phi_{i\mu}$, out of which we select singlet CSFs 
($\mu$ such that $s_\mu=0$). Finally, we project the Hamiltonian matrix in the singlet subspace, constructing the operator
\begin{equation}
\tilde{H} = \left[ \, \sum_{\mu=0}^{K-1} | \phi_\mu \rangle \langle \phi_\mu | \, \right]
\hat{H} 
\left[ \, \sum_{\nu=0}^{K-1} | \phi_\nu \rangle \langle \phi_\nu | \, \right]
\quad.
\end{equation}
Where $K$ is the number of singlet CSFs in the target irrep. In some cases, $K=2^{n_q}$ for some number of qubits $n_q$, in which case $\tilde{H}$ can be represented as a qubit operator by expanding it on the Hilbert-Schmidt basis,
\begin{equation}
\tilde{H} = \sum_{\vett{v} \, \vett{w}} \sigma_{\vett{v} \, \vett{w}} \; \frac{ \mbox{Tr}\big[ \sigma_{\vett{v} \, \vett{w}} \tilde{H} \big] }{2^{n_q} } \quad.
\end{equation}
In general, $2^{n_q-1} < K < 2^{n_q}$ for some integer $n_q$, and $\tilde{H}$ can be 
represented as a qubit operator in at least two ways:
\begin{enumerate}
\item by ``trimming'' the matrix $\langle \phi_\mu | \hat{H} | \phi_\nu \rangle$ removing a subset of $K-2^{n_q-1}$ CSFs. 
In the present work, we chose to compute the ground-state wavefunction of $\tilde{H}$,
$| \psi_0 \rangle = \sum_\mu c_\mu | \phi_\mu \rangle$ and to retain the $2^{n_q-1}$ CSFs with the largest coefficients in magnitude $|c_\mu|$.
\item by introducing a set of $2^{n_q}-K$ additional basis functions, 
called unphysical basis function, ``padding'' the matrix $\tilde{H}_{\mu\nu}$ 
to produce a $2^{n_q} \times 2^{n_q}$ matrix, and introducing a projector 
onto the span of physical basis functions,
\begin{equation}
\label{eq:p_php}
J_{\mu\nu} = 
\left(
\begin{array}{c|c}
\tilde{H} & {\bf{0}} \\
\hline
{\bf{0}} & \lambda \mathbbm{1} \\
\end{array} 
\right)
\quad,\quad
\Pi_{\mu\nu} =
\left(
\begin{array}{c|c}
\mathbbm{1} & {\bf{0}} \\
\hline
{\bf{0}} & {\bf{0}} \\
\end{array} 
\right)
\quad .
\end{equation}
\end{enumerate}
In this work, we chose $\lambda = 10^4 \, \mathrm{E_h}$. When first-quantization calculations are carried out within the ``padding'' scheme, qubit wavefunctions may have a component outside the ``physical'' subspace (the image of the projector $\hat \Pi$).
To remove such unphysical wavefunctions, two schemes can be deployed:
\begin{enumerate}
\item Variation-after-projection (VAP) \cite{scuseria2011projected,jimenez2012projected}. The following cost function is optimized,
\begin{equation}
\label{eq:vap_cost_function}
E_{\mathrm{vap}}(\bgreek{\theta}) = \frac{ \langle \Psi(\bgreek{\theta}) | \hat \Pi \hat{J} \hat \Pi | \Psi(\bgreek{\theta}) \rangle }{ \langle \Psi(\bgreek{\theta}) | \hat\Pi | \Psi(\bgreek{\theta}) \rangle } \quad ,
\end{equation}
i.e. the wavefunction $| \Psi(\bgreek{\theta}) \rangle$ is first projected in the physical subspace, and the energy is then optimized.
\item Projection-after-variation (PAV) \cite{mcdouall1989analytical,hratchian2013communication,thompson2015second}. The cost function
\begin{equation}
E_{\mathrm{pav}}(\bgreek{\theta}) = \langle \Psi(\bgreek{\theta}) | \hat{J} | \Psi(\bgreek{\theta}) \rangle
\end{equation}
is optimized, and the energy is then computed with Eq. \eqref{eq:vap_cost_function},
i.e. over a projected wavefunction.
\end{enumerate}
In this study, we employ the VAP scheme, as it yields more accurate results than the PAV  scheme \cite{scuseria2011projected,jimenez2012projected}. It should be noted that both schemes may yield very inaccurate results when the qubit wavefunction lies predominantly outside the physical subspace, i.e. when $P = \langle \Psi(\bgreek{\theta}) | \hat\Pi | \Psi(\bgreek{\theta}) \rangle \ll 1$.

\subsection{Target variational Ans\"{a}tze}

In this study, we elected to explore the behavior of the variational quantum eigensolver 
(VQE) algorithm \cite{peruzzo2014variational}, due to its widespread adoption in the community.
This method defines a set of Ansatz states approximating the ground-state of a target Hamiltonian, of the form 
\begin{equation}
| \Psi( \bgreek{\theta} ) \rangle = \hat{U}( \bgreek{\theta} ) | \Phi_0 \rangle
\quad,
\end{equation}
where $\bgreek{\theta} \in [0,2\pi)^{n_\theta}$ is an array of angles that define a 
parametrized quantum circuit $\hat{U}( \bgreek{\theta} )$, applied to a register of 
$n_q$ qubits prepared in a standard initial state $| \Phi_0 \rangle$, for example $| \Phi_0 \rangle = | 0 \rangle^{\otimes n_q}$.
The best approximation to the ground-state in the set of Ansatz states is found by 
minimizing the energy 
\begin{equation}
E_{\mathrm{VQE}}( \bgreek{\theta} ) = \langle \Psi( \bgreek{\theta} ) | \hat{H} | \Psi( \bgreek{\theta} ) \rangle
\end{equation}
as a function of the parameters $\bgreek{\theta} $ using a classical optimization 
algorithm. Once the parameters are optimized, the auxiliary observables 
Eq.~\eqref{eq:auxiliary} can be measured, yielding results
\begin{equation}
X_{\mathrm{VQE}}( \bgreek{\theta} ) 
= 
\langle \Psi( \bgreek{\theta} ) | \hat{X} | \Psi( \bgreek{\theta} ) \rangle
\;,\;
\hat{X} = \hat{N}_e, \hat{S}_z, \hat{S}^2
\;.
\end{equation}
This algorithmic workflow, termed variational quantum eigensolver (VQE) in the quantum 
simulation literature, is a technique for ground-state wavefunction approximation. 
Its accuracy, computational cost, and potential pitfalls 
%discussed in Section \ref{sec:metrics}, 
are determined by the details of the circuit $\hat{U}( \bgreek{\theta} )$. 
We now outline the three hardware-efficient Ans\"{a}tze studied in this work.

\subsubsection{Hardware-efficient $\ry$}

\begin{figure}[h!]
\includegraphics[width=\textwidth]{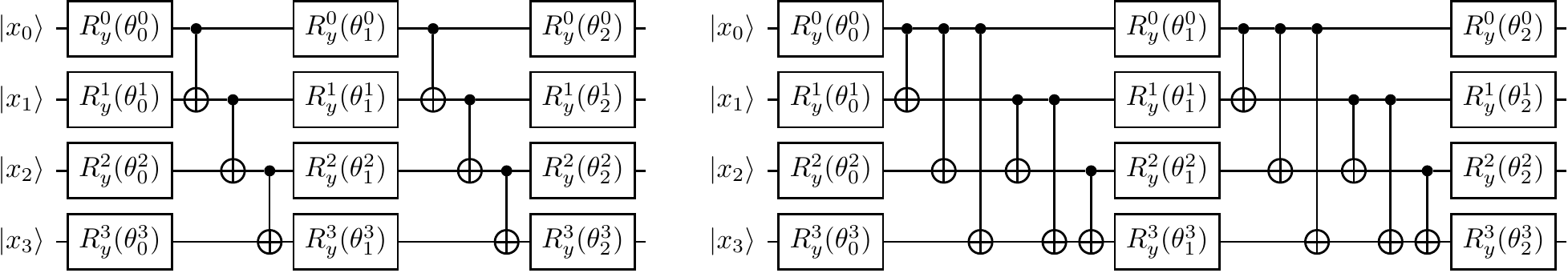}
\caption{Quantum circuits implementing the linear- and full-connectivity $\ry$ Ansatz, 
with $n_q=4$ qubits and $n_l=2$ layers (left, right respectively). The illustrated circuits have a computational basis state or a bit-string, 
i.e. a tensor product of the form $| {\bf{x}} \rangle = \otimes_{k=0}^{n_q-1} |x_k \rangle$ with $x_k \in \{0,1\}$, 
as initial state. $|0 \rangle$ is the ground-state of a single qubit, and $|1\rangle$ can be prepared starting from a qubit in $|0 \rangle$ by applying a single-qubit rotation of an angle $\pi$ around the $x$ axis of the Bloch sphere, $R_x(\pi) = \exp(-i \pi X /2)$.
}
\label{figure:ry}
\end{figure} 

Hardware-efficient Ans\"{a}tze are families of wavefunctions designed with the primary goal to be compatible with the budget of near-term quantum hardware 
in terms of available qubits, connectivity, native gates, and circuit depth.
An example is the following $\ry$ Ansatz \cite{kandala2017hardware},
\begin{equation}
| \Psi(\theta) \rangle = 
\prod_{k=1}^{n_l}
\left[ \mathcal{R}(\theta^{n_q-1}_{k} \dots \theta^0_{k})
\prod_{ij\in C} \mathsf{c}_i \mathsf{NOT}_j
\right]
\mathcal{R}(\theta^{n_q-1}_{0} \dots \theta^0_{0}) | \Phi_0 \rangle
\quad,\quad
\mathcal{R}(\theta^{n_q-1} \dots \theta^0) = \prod_{i=0}^{n_q-1} R_{y,i}(\theta^i) 
\quad,
\end{equation}
where $| \Phi_0 \rangle$ is an initial wavefunction (here, the restricted 
closed-shell Hartree-Fock state), $n_q$ is the number of qubits, $R_{y,i}(\theta) 
= \mbox{exp}(-i \theta Y_i/2)$ is a $Y$-rotation (i.e. a rotation of an angle $\theta$ around the $y$ axis of qubit $i$'s Bloch sphere),
$\mathsf{c}_i \mathsf{NOT}_j$ is a CNOT gate with control qubit $i$ and target qubit $j$, and $n_l$ is an integer denoting the number of times a layer of entangling gates
followed by a layer of $Y$-rotations is repeated. 

In this study, we focused on the {\em{linear-connectivity}} and {\em{full-connectivity}} $\ry$ Ans\"{a}tze, given by
\begin{equation}
\label{eq:ry_connectivity}
C_{\mathrm{linear}} = \{ (i,i+1) \;,\; i = 0 \dots n_q-2 \}
\quad,\quad
C_{\mathrm{full}} = \{ (i,j) \;,\; i,j = 0 \dots n_q-1 \;,\; i \leq j \}
%\quad. 
\end{equation}
respectively.
Examples of linear- and full-connectivity $\ry$ Ans\"{a}tze are shown in Figure \ref{figure:ry}.
We note that other choices of connectivity are possible and that Eq.~\eqref{eq:ry_connectivity} exemplifies two opposite regimes.

\subsubsection{Cascade}

\begin{figure}[h!]
\includegraphics[width=0.78\textwidth]{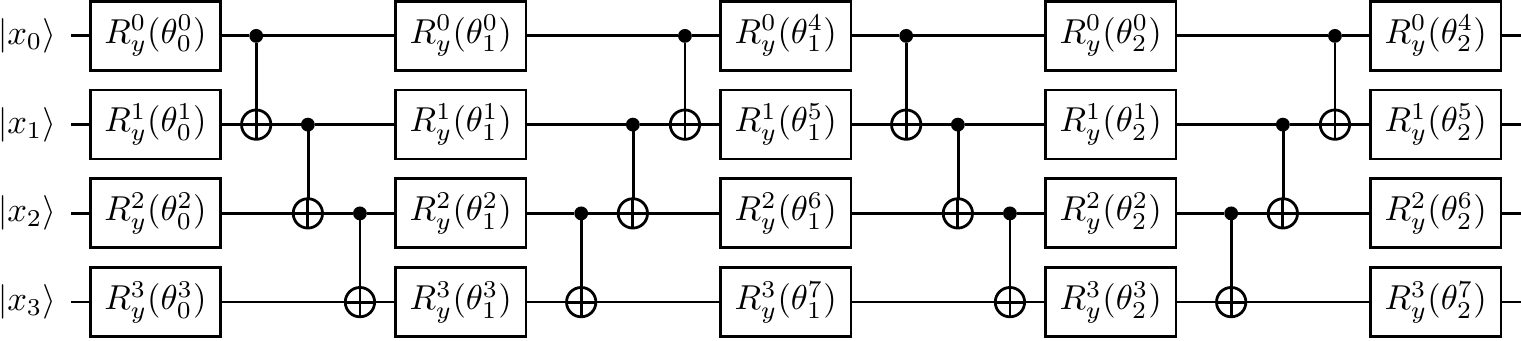}
\caption{Quantum circuit implementing the cascade Ansatz, with $n_q=4$ qubits and $n_l=2$ layers.}
\label{figure:cascade}
\end{figure} 

In standard qubit representations, the Hartree-Fock state is mapped onto a computational basis state, or a bit-string, $| \Phi_0 \rangle = \otimes_{i=0}^{n_q-1} |x_i \rangle = | {\bf{x}} \rangle$.
When applied to an initial state of the form $| \Phi_0 \rangle = | {\bf{x}} \rangle$, the $\ry$ Ansatz with $\bgreek{\theta} = {\bf{0}}$ returns a bitstring $| {\bf{x}}^\prime \rangle$ potentially with
different properties than $| {\bf{x}} \rangle$, e.g. higher energy, higher/lower particle number. 
In some situations, it is desirable to employ Ans\"{a}tze with the property that 
$\hat{U}({\bf{0}}) = \mathbbm{1}$, so that the starting point of the optimization  
is the energy of the initial state (usually the Hartree-Fock state).
An example is the following Ansatz, that we term ``cascade'', illustrated in Figure \ref{figure:cascade} and defined by the equation
\begin{equation}
| \Psi(\bgreek{\theta}) \rangle 
= \prod_{k=1}^{n_l} \left[ \mathcal{R}(\theta_k^{2 n_q-1} \dots \theta_k^{n_q}) C^\dagger
\mathcal{R}(\theta_k^{n_q-1} \dots \theta_k^0) C \right]
\mathcal{R}(\theta_0^{n_q-1} \dots \theta_0^0)| \Psi_0 \rangle
\quad,
\end{equation}
where
%\begin{equation}
$C = \mathsf{c}_{n_q-2}\mathsf{NOT}_{n_q-1} \dots \mathsf{c}_0\mathsf{NOT}_1$
is a ladder of CNOT gates applied to adjacent qubits, and $\mathcal{R}$ a product of $Y$-rotations, exactly as in the $R_y$ Ansatz.
As seen, the presence of $C$ and $C^\dagger$ ensures that $\hat{U}({\bf{0}}) = \mathbbm{1}$.
Furthermore, as in other Ans\"{a}tze \cite{farhi2014quantum,ryabinkin2018qubit}, given a wavefunction with $n_l$ repetitions and optimized parameters $\bgreek{\theta}_{n_l}$, 
a wavefunction with $n_l+1$ repetitions can be initialized from the parameter configuration $\bgreek{\theta}_{n_l+1} = ( \bgreek{\theta}_{n_l} , {\bf{0}} )$ 
so that the VQE energy, unlike that from the $R_y$ Ansatz, decreases monotonically with $n_l$.

\subsubsection{Unitary coupled-cluster with singles and doubles}

To assess the accuracy of the hardware-efficient Ans\"{a}tze and highlight their potential pitfalls, in this work we will compare
them against full configuration interaction (FCI) and the quantum unitary coupled-cluster (q-UCC) method.

The quantum unitary coupled-cluster (q-UCC) method is based on the exponential 
Ansatz, i.e., the exact wavefunction (within a chosen one-
electron basis) is written as \cite{bartlett1989alternative,Moll_2018,romero2018strategies,Cooper2010,harsha2018difference,evangelista2019exact}
\begin{equation}
\label{eq:qucc}
| \Psi_{\mathrm{gs}} \rangle = e^{ \hat{T} - \hat{T}^\dagger } | \Phi_0 \rangle
\quad,
\end{equation}
where $\Phi_0$ is typically an independent-particle function (here, the Hartree-Fock state) 
and $\hat{T}$ is a cluster operator which, at the singles and doubles level 
(q-UCCSD) is truncated to 
\begin{equation}
\label{eq:unrestricted_quccsd}
\hat{T} = \hat{T}_1 + \hat{T}_2 \quad,\quad
\hat{T}_1 = \sum_{\substack{ai \\ \sigma}} t^{a \sigma}_{i \sigma} \crt{a \sigma} \dst{i \sigma} \quad,\quad
\hat{T}_2 = \frac{1}{4} \sum_{\substack{abij \\ \sigma\tau}} t^{a\sigma \; b\tau}_{i\sigma \; j\tau} \crt{a \sigma} \crt{b \tau} \dst{j \tau} \dst{i \sigma} \quad,
\end{equation}
where $t^{a \sigma}_{i \sigma}$ and $t^{a\sigma \; b\tau}_{i\sigma \; j\tau}$ are a set of unknown cluster coefficients. 
We have adopted the convention that $i,j,k,l$ and $a,b,c,d$ refer to occupied 
and unoccupied orbitals in the reference state $\Phi_0$ respectively.

Eq.~\eqref{eq:unrestricted_quccsd}, 
presented in previous literature \cite{barkoutsos2018quantum} along with the corresponding quantum circuit, is an unrestricted Ansatz, which is not guaranteed to yield an eigenfunction of total spin.
If a closed-shell spin-adapted formalism is used, then, following Paldus \cite{paldus1977correlation} and Scuseria {\em{et al}} \cite{scuseria1987closed,szalay1997spin}, one needs 
\begin{equation}
\label{eq:restricted_quccsd_s}
t^{a \alpha}_{i \alpha} = t^{a \beta}_{i \beta} = t^a_i \quad,
\end{equation}
since the two other possible spin combinations are vanishing. 
Only 6 of the 16 possible spin combinations for the two-electron amplitudes 
are non-zero, namely
\begin{equation}
\label{eq:restricted_quccsd_d}
t^{a \alpha \; b \alpha}_{i \alpha \; j \alpha} = t^{a \beta \; b \beta}_{i \beta\; j\beta} = \hat{t}^{ab}_{ij} 
\quad,\quad
t^{a \alpha \; b \beta}_{i \alpha \; j \beta} = t^{a \beta \; b \alpha}_{i \beta\; j\alpha} = \tilde{t}^{ab}_{ij} 
\quad,\quad
t^{a \beta \; b \alpha}_{i \alpha \; j \beta} = t^{a \alpha \; b \beta}_{i \beta\; j\alpha} = \overline{t}^{ab}_{ij} 
\quad .
\end{equation}
Using the relations
\begin{equation}
\hat{t}^{ab}_{ij} = \tilde{t}^{ab}_{ij} + \overline{t}^{ab}_{ij}
\quad,\quad
\tilde{t}^{ab}_{ij} = - \overline{t}^{ab}_{ji} = - \overline{t}^{ba}_{ij} = \tilde{t}^{ba}_{ji}
\quad,
\end{equation}
one is left with only one set of independent two-electron amplitudes $\tilde{t}^{ab}_{ij}$ with 
$a \leq b$,  $i \leq j$ and
$(ai) \leq (bj)$. In addition to reducing the 
number of variational parameters, the use of a closed-shell spin-adapted formalism ensures that the cluster and total spin
operators commute, $[ \hat{T} , \hat{S}^2 ] = 0$.

{Implementing the exponential in Eq.~\eqref{eq:qucc} on a quantum computer is not straightforward. Here, we illustrate the effect of operator order and Trotter approximation on the accuracy and spin symmetry-breaking properties of q-UCCSD \cite{grimsley2019trotterized,tsuchimochi2020spin}. 
It is important to remark that the Trotterized form of q-UCC is not exact, and is 
order-dependent \cite{evangelista2019exact}. Order independence and exactness can be restored by Lie algebraic closure of the generator set \cite{izmaylov2020order}.
}

\subsection{Properties studied}
\label{sec:metrics}

{
The quantum information science community has already identified the variational optimization of parameters as a bottleneck of hardware-efficient (and other) Ans\"{a}tze, due to the presence of barren plateaus \cite{mcclean2018barren}.
In our study, we illustrate that the limitations of hardware-efficient Ans\"{a}tze in their application to ES are even more profound and nuanced.
Hardware-efficient Ansätze 
can break Hamiltonian symmetries, e.g. yielding wavefunctions that do not have a well-defined spin or particle number.
Furthermore, they 
can lead to potential energy curves that are not differentiable with respect to nuclear coordinates, leading to ill-defined forces.
Finally, even supplying adequate initial guesses for the parameter optimization may be challenging. These pitfalls are so pronounced that they are exhibited even by closed-shell molecular species at the STO-6G level of theory. Furthermore, their interplay is rather subtle. After illustrating the limitations of hardware-efficient Ans\"{a}tze for a set of second-quantization problems,  we perform a comparative analysis of data, to illustrate the interplay between non-differentiability, symmetry-breaking, and challenging optimization.
}

For the small molecules studied here, the exact solution of the Schr\"{o}dinger equation can be accessed using FCI. A natural metric \cite{LeBlanc_PRX_2015,Zheng_Science_2017,Motta_PRX_2017,williams2020direct} to assess the accuracy of a variational simulation is therefore the difference between the computed and exact ground-state energy (in the same one-electron basis), 
\begin{equation}
\label{eq:delta_E}
\Delta E = E_{\mathrm{VQE}}(\bgreek{\theta}) - E_{\mathrm{FCI}}
\quad .
\end{equation}
{The computation of total energies and deviations from FCI across dissociation profiles also allows exploration of the conditions under which hardware-efficient Ans\"{a}tze yield non-differentiable potential energy curves.}

As mentioned at the beginning of this Section, symmetry-breaking is a pitfall of hardware-efficient Ans\"{a}tze. The ES Hamiltonian has several symmetries, i.e. operators $\hat{X}$ such that $[\hat{H},\hat{X}]=0$.
In the presence of Hamiltonian symmetries, the search for Hamiltonian eigenfunctions has to be restricted to eigenspaces of $\hat{X}$ with known eigenvalues.
Relevant Hamiltonian symmetries are the auxiliary operators Eq.~\eqref{eq:auxiliary}
and, for the molecules studied in the present work, the ground-state should lie in the eigenspaces of $\hat{S}_z$ and $\hat{S}^2$ with eigenvalue zero,
and of $\hat{N}$ with eigenvalues $2$ for LiH, $4$ for BH and BeH$_2$, $6$ for H$_2$O and $8$ for HF.
To assess whether a variational simulation leads to symmetry-breaking, and to quantify its extent, we computed the differences
\begin{equation}
\Delta N = N_{\mathrm{VQE}}(\bgreek{\theta}) - N_{\mathrm{FCI}} 
\quad,\quad
\Delta S_z = S_{z,\mathrm{VQE}}(\bgreek{\theta}) - S_{z,\mathrm{FCI}}
\quad,\quad
\Delta S^2 = S^2_{\mathrm{VQE}}(\bgreek{\theta}) - S^2_{\mathrm{FCI}}
\quad,\quad
\end{equation}
between the calculated and exact ground-state particle number, spin-$z$, and total spin.

\section{Results and Discussion}
\label{sec:results}

The overall strategy for the calculations performed in this work involved initial pre-processing by the classical quantum chemistry code PySCF \cite{sun2018pyscf,sun2020recent} (as detailed in Section II of the SI)
to generate optimized Hartree-Fock orbitals, Hamiltonian coefficients, and FCI energies and properties, before performing computations with quantum simulators.
The restricted Hartree-Fock (RHF) singlet state was chosen as the initial state for all of the calculations described here, 
since experience has indicated that this state is a good choice for various chemical problems \cite{romero2018strategies}.
Details about the molecular orbitals and the FCI natural orbitals are in Section II of the SI.

Having selected a set of single-electron orbitals for each studied species, we performed VQE computations with quantum simulators. 
We used IBM's open-source Python library for quantum computing, Qiskit \cite{aleksandrowicz2019qiskit}. 
Qiskit provides tools for various tasks, for example creating quantum circuits and performing quantum simulations. 
In particular, it contains an implementation of the VQE algorithm.

Following the VQE protocol, we then minimized the expectation value of the Hamiltonian with respect to the parameters of the circuit, $\bgreek{\theta}$. To carry out the minimization we employed
the classical optimization methods L$\_$BFGS$\_$B and conjugate gradient \cite{zhu1997algorithm,morales2011remark}.
In many cases, to determine the lowest energy along the dissociation path, we performed multiple VQE calculations using different initial parameters.
Once the VQE was complete, we obtained the optimized variational form and the estimate for the ground-state energy. 
In addition, we measured the set of auxiliary operators, see Eq.~\eqref{eq:auxiliary} to investigate the degree of symmetry breaking.

\subsection{Second-quantization simulations}
\label{sec:results_2nd}

\begin{figure}[h!]
\includegraphics[width=0.53\textwidth]{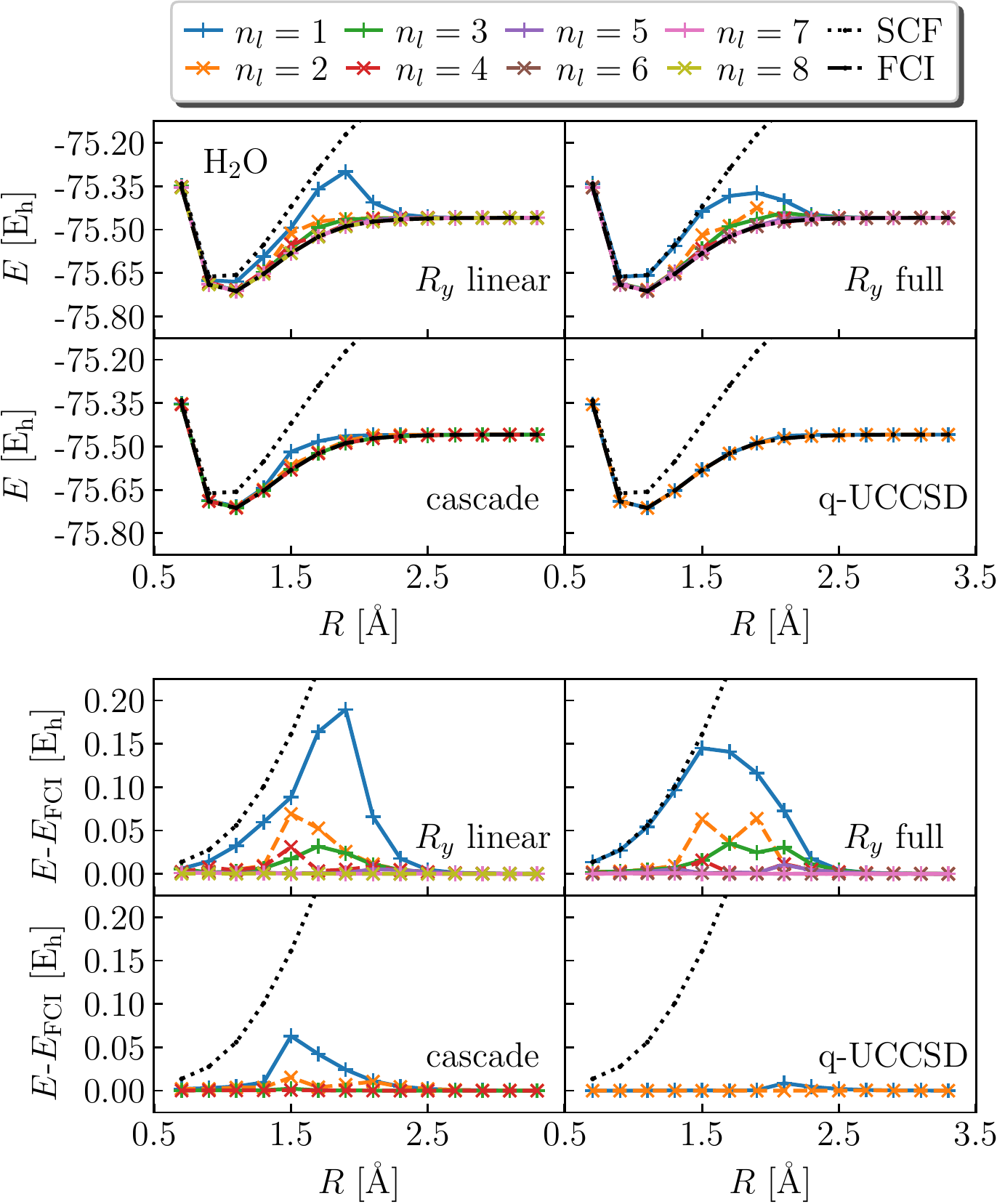}
\caption{Total energy (top chart) and deviation between computed and exact total energy  (bottom chart) using the R$_y$ (with linear- and full-connectivity), cascade, and q-UCCSD Ans\"{a}tze (left to right), 
for the H$_2$O molecule at the STO-6G level.
}
\label{figure:second_h2o_E}
\end{figure}

\begin{figure}[h!]
\includegraphics[width=0.53\textwidth]{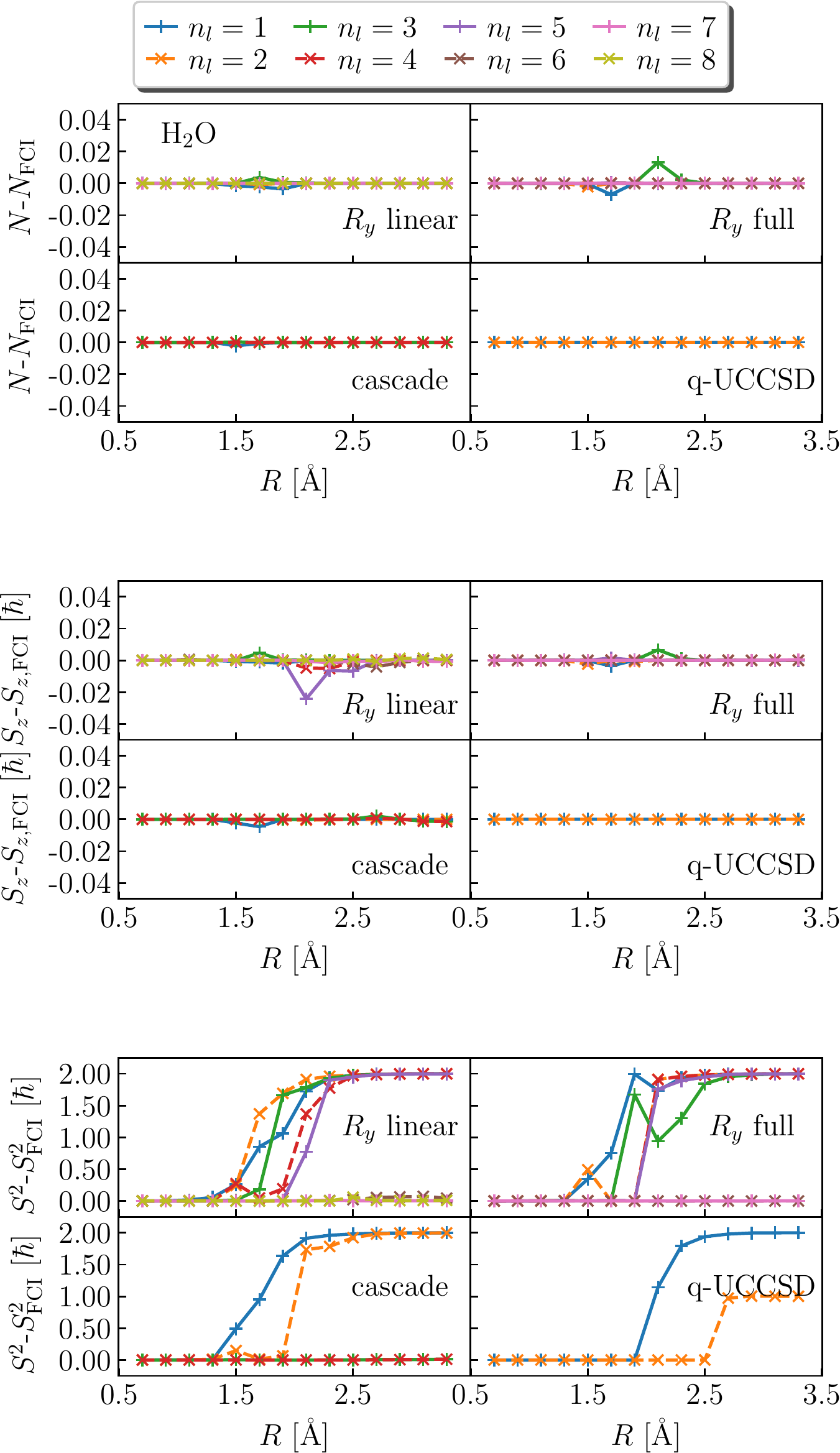}
\caption{Deviation between computed and exact electron number, total spin, and spin-$z$ (top to bottom) using the R$_y$ (with linear- and full-connectivity), cascade, and q-UCCSD Ans\"{a}tze (left to right), 
for the H$_2$O molecule at the STO-6G level.
}
\label{figure:second_h2o_P}
\end{figure}

{
In Figures \ref{figure:second_h2o_E} and \ref{figure:second_h2o_P}, we show the ground-state energy and auxiliary properties (as indicators of symmetry-breaking) of the H$_2$O molecule within second-quantization encoding. 
}

For the bond lengths and Ans\"{a}tze studied here, energies typically decrease monotonically with number of Ansatz layers $n_l$.
As naturally expected, hardware-efficient Ans\"{a}tze are outperformed by q-UCCSD, since the latter is constructed starting from the natural excitations of the system (i.e. electronic transitions from occupied to virtual orbitals), rather than from application-agnostic circuits.

Interestingly, the linear-connectivity $\ry$ Ansatz typically outperforms its full-connectivity counterpart, achieving more accurate results with fewer quantum resources.
Furthermore, a cascade Ansatz with $n_l$ layers performs approximately as well as a linear-connectivity $\ry$ Ansatz with $2 n_l$ layers.

The most challenging regime is the intermediate dissociation region ($R \simeq 2$-$3 \, \ang$), where the energies from hardware-efficient Ans\"{a}tze are more than 100 milliHartree above the FCI value and clearly display cusps.
Since calculations were carefully converged, these cusps are likely not a product of 
unconverged parameters, but rather a genuine limitation of the underlying Ans\"{a}tze,
which stands to impact the computation of energy gradients and lead to nonphysical forces.

It should be noted that an inaccurate performance by hardware-efficient Ans\"{a}tze and q-UCCSD is accompanied by a pronounced breaking of the spin symmetry: the total spin of low-depth $\ry$ Ans\"{a}tze evolves from $S^2 = 0$ at short $R$ towards a mixture of singlet and triplet at large $R$.
Ansatz inaccuracy and symmetry-breaking, however, are not perfectly correlated. The largest deviations between VQE and FCI energies occur in the intermediate dissociation region, where symmetry-breaking starts to develop due to a larger static correlation; 
on the other hand, spin symmetry-breaking is most pronounced at dissociation ($R \geq 3 \ang$), where the VQE energy is in better agreement with FCI. In the dissociation limit, multiple excited states with higher spin have the same energy as the singlet. We attribute the combination of pronounced symmetry-breaking and high accuracy of energies in the dissociation limits to the fact that simulations target high-spin states nearly degenerate with the singlet ground state.
{Note that symmetry-breaking also occurs in q-UCCSD calculations (vide infra).

The pitfalls of hardware-efficient Ans\"{a}tze illustrated in Figures \ref{figure:second_h2o_E} and \ref{figure:second_h2o_P} are especially pronounced for the H$_2$O molecule
where two bonds are broken simultaneously. However, they are general features: for example, as listed in Table \ref{table:second_quantization} and shown in Section V of the SI, they affect other molecules, albeit to different extents.
BeH$_2$ exhibits pronounced symmetry-breaking as highlighted by $S^2$, $S_z$, and $N_e$ (also in this case two bonds are broken). 
HF and BH are also affected by symmetry-breaking and non-differentiability of potential energy curves. Even LiH, a two-electron system that has an exact solution from any classical method with single and double excitations (e.g. CCSD), is affected by symmetry-breaking. 

\begin{table}[h!]
\begin{tabular}{cccccc}
\hline\hline
molecule & $\Delta N_e$ & $\Delta S_z [\hbar]$ & $\Delta S^2 [\hbar]$ & $\Delta E [\mathrm{E_h}]$ & cusps \\
\hline
H$_2$O & 0.013 & 0.024 & 2.00 & 0.189 & yes \\ 
BeH$_2$ & 0.012 & 0.033 & 2.00 & 0.201 & yes \\
BH & 0.002 & 0.001 & 2.00 & 0.083 & yes \\
HF & 0.001 & 0.000 & 1.00 & 0.069 & yes \\
LiH & 0.002 & 0.001 & 1.00 & 0.018 & no \\
\hline\hline
\end{tabular}
\caption{Maximum absolute differences between computed and exact particle number, total spin-$z$, total spin, and ground-state energy for all the molecules in this study. The maximum is identified over all studied bond lengths, Ans\"{a}tze, and numbers of layers (see Section V of the SI). The last column indicates whether potential energy curves by hardware-efficient Ans\"{a}tze display cusps.}
\label{table:second_quantization}
\end{table}
}

\subsubsection{Optimization of cascade and $R_y$ Ans\"{a}tze}

\begin{figure}[h!]
\includegraphics[width=0.75\textwidth]{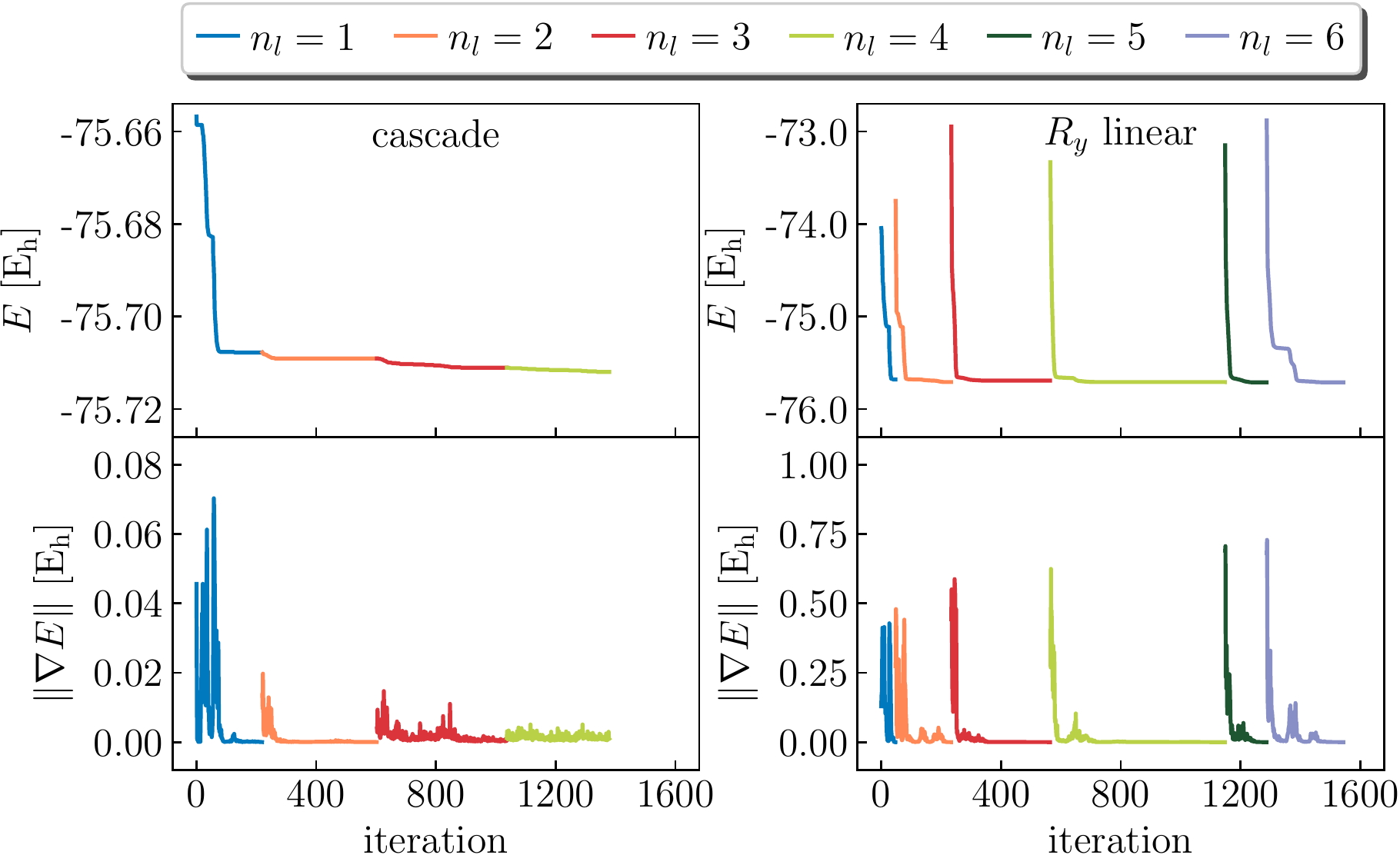}
\caption{Energy (top) and  energy gradient (bottom) during the optimization of cascade (left) and linear-connectivity $\ry$ (right) for the H$_2$O molecule at $R=1.1$ $\mathrm{\AA}$.
}
\label{figure:optimization}
\end{figure}

In Figure \ref{figure:optimization} we show the optimization of cascade and $\ry$-linear Ans\"{a}tze for H$_2$O near the equilibrium geometry. Optimizations are carried out using BFGS. In both cases, the $n_l=1$ calculation is initialized from $\bgreek{\theta}={\bf{0}}$. Given a set of optimized
parameters for $n_l$, the $n_l+1$ calculation is initialized initialing the extra parameters with zeros.
As seen, cascade features a monotonically decreasing energy with $n_l$, whereas $\ry$ optimizations do not. However, the total number of iterations is roughly the same for both Ans\"{a}tze.

We note that, as $n_l$ increases, optimizations become noise-dominated (see e.g. small gradient of cascade for $n_l=4$). This behavior makes very challenging to simultaneously converge energy, and properties within $10^{-2}$, across dissociation.

\subsubsection{Symmetry-breaking in q-UCCSD calculations}
\label{sec:q-UCCSD}

\begin{figure}[h!]
\includegraphics[width=0.7\textwidth]{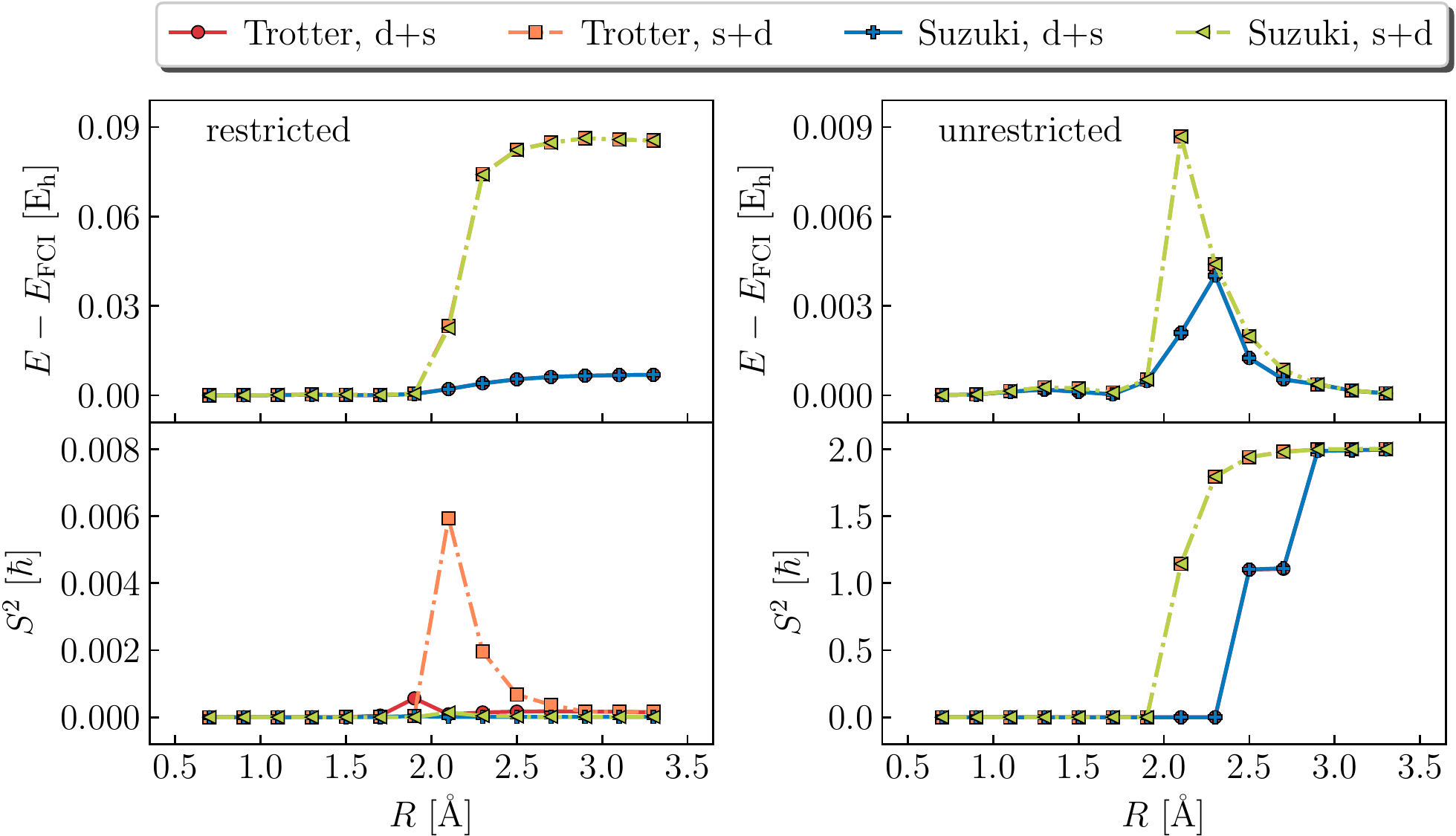}
\caption{Deviation between computed and exact energy (top) and total spin (bottom), for the restricted closed-shell (left) and unrestricted (right) q-UCCSD Ans\"{a}tze with $n_l=1$ layers, for the H$_2$O molecule at the STO-6G level, using Trotter and Suzuki approximations (warm, cold colors) and with singles followed by doubles and doubles followed by singles (light, dark colors).}
\label{figure:quccsd_reps_1}
\end{figure} 

{The q-UCCSD calculations reported in Figures \ref{figure:second_h2o_E} and \ref{figure:second_h2o_P} are spin-unrestricted, employ the primitive Trotter approximation to exponentiate the cluster operator, with exponentials of singles before exponentials of doubles. This choice led to a spin-contaminated solution for H$_2$O at large $R$, see Fig.~\ref{figure:second_h2o_P}.
It is well-known that the Trotterized form of q-UCC is not exact and 
order-dependent \cite{evangelista2019exact,izmaylov2020order}. The details of the algorithmic implementation of a Trotterized q-UCC have an effect on its ability to produce eigenstates of total spin.} In Figure~\ref{figure:quccsd_reps_1} we compare the energies and total spins of restricted closed-shell and unrestricted q-UCCSD, approximated with primitive Trotter and second-order Suzuki product formulas and $n_l=1$ layers,
with singles followed by doubles and doubles followed by singles, for the H$_2$O molecule at the STO-6G level.
The quantum circuits used to implement the q-UCCSD Ansatz in a Jordan-Wigner representation are illustrated in Section III of the SI.

As seen, the unrestricted Ansatz yields lower energies at large $R$, 
albeit with a significant spin contamination (a phenomenon analogous to 
the Coulson-Fischer \cite{coulson1949xxxiv} instability of mean-field theories).
The restricted Ansatz, though leading to higher energies, features a
modest spin contamination entirely caused by the Trotter approximation,
\begin{equation}
e^{\hat{T} - \hat{T}^\dagger} 
=
e^{ \sum_\mu t_\mu (\hat{O}_\mu-\hat{O}_\mu^\dagger)}
\simeq 
\left[ 
\prod_{\mu} e^{ \frac{t_\mu (\hat{O}_\mu-\hat{O}_\mu^\dagger) }{n_l} }
\right]^{n_l}
\quad,\quad
e^{ \frac{t_\mu (\hat{O}_\mu-\hat{O}_\mu^\dagger) }{n_l} }
\simeq 
\prod_{i}
e^{ \frac{t_\mu c_{i\mu}}{n_l} \sigma_{{\bf{v}}_i {\bf{w}}_i} }
\;,
\end{equation}
where $n_l$ is the number of Trotter steps and $\sum_i c_{i\mu} \sigma_{{\bf{v}}_i {\bf{w}}_i}$ is a qubit representation of $\hat{O}_\mu-\hat{O}_\mu^\dagger$ {(see Section III of the SI for details)}.
Symmetry-breaking phenomena are milder when a second-order Suzuki 
product formula is used, and deviations with FCI are more modest
when doubles are followed by singles. This is not unexpected \cite{grimsley2019trotterized}, since the 
mean-field reference state only couples with doubles due to Brillouin's theorem.

In particular, a proper singlet wavefunction can be obtained without any accuracy loss by using a restricted closed-shell implementation of q-UCCSD, Suzuki approximation, doubles before singles and $n_l=2$ layers.
{We remark that q-UCCSD does not feature barren plateaus in the parameter optimization; this is a simple manifestation of the fact that symmetry-breaking and difficult parameter optimization are not perfectly correlated.}

\subsection{First-quantization simulations}

Second-quantization make use of the full Fock space of electrons in $m$ spatial orbitals. 
As observed in the previous Section, this choice can lead to various symmetry-breaking phenomena, 
due to qubit wavefunctions having a component outside the target subspace (e.g. $N_e = N_\alpha + N_\beta$, $S_z = (N_\alpha - N_\beta)/2$,
and $S^2 = S_z (S_z+1) \hbar$).

{A possible strategy to remedy such a pitfall of hardware-efficient data is to rely on a first-quantization encoding, which automatically gives access to eigenfunctions of particle number, spin, and molecular orbital point-group symmetry.}

\begin{figure}[h!]
\includegraphics[width=0.85\textwidth]{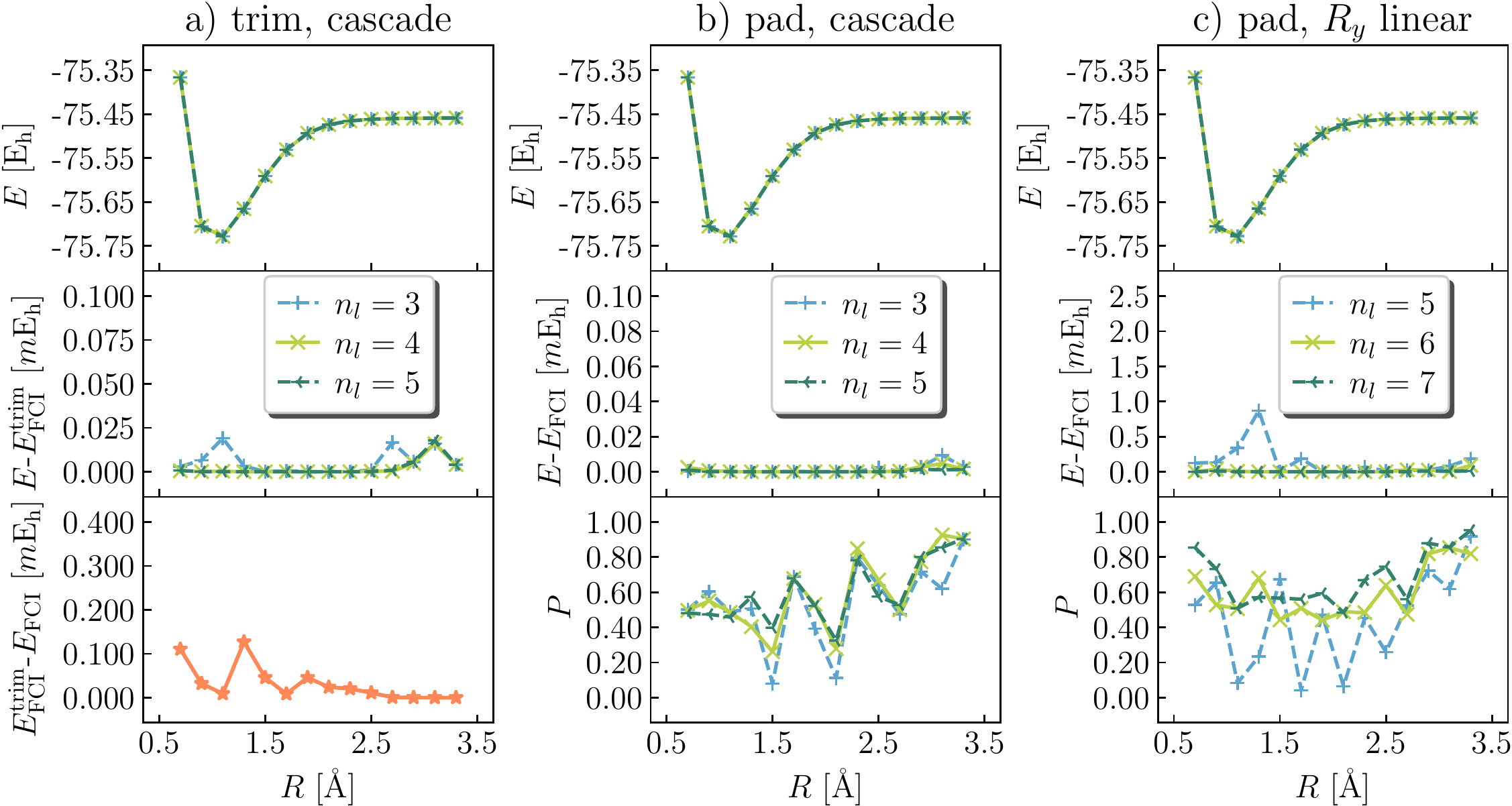}
\caption{Left (a): total energy, deviation between computed and exact total energy using a first-quantization encoding and a trimming procedure for qubit reduction, and energy error from the trimming procedure (top to bottom) for the H$_2$O molecule at the STO-6G level using the cascade Ansatz.
Middle (b): total energy, deviation between computed and exact energy, and squared-norm of the physical component of the wavefunction (top to bottom) for the H$_2$O molecule at the STO-6G level using a first-quantization encoding with padding and a cascade Ansatz optimized with the variation-after-projection scheme.
Right (c): same as the middle panel, with a linear-connectivity $\ry$ Ansatz.
}
\label{figure:first_quantization_results}
\end{figure}

In Figure \ref{figure:first_quantization_results}a,b we computed ground-state potential energy curves for H$_2$O using the cascade Ansatz
and the ``trimming'' and ``padding'' schemes. In Figure \ref{figure:first_quantization_results}c, we use instead the linear-connectivity $\ry$ Ansatz.
In this framework, cascade simulations with $n_l=3$ or greater agree with the exact ground-state energy of the trimmed Hamiltonian within 25 microHartree. However, the trimming itself leads to a deviation between the FCI energy of the trimmed and that of the original first-quantized Hamiltonian, which is up to 0.1 milliHartree.
We also remark that all potential energy curves exhibit cusps.
The squared-norm $P$ of the physical component of the wavefunction reaches values as small as $2 \cdot 10^{-1}$ and $1 \cdot 10^{-2}$ for cascade and $\ry$ respectively. Such a phenomenon indicates that, for both Ans\"{a}tze, lowering the energy is accompanied by leakage of the wavefunction outside the physical subspace.

{The first-quantization data presented in this Section indicate that discontinuities in the potential energy curves can be observed in absence of symmetry-breaking,  hence the two phenomena are not perfectly correlated.}

\subsection{Computational cost}

\begin{table}[h!]
\begin{tabular}{cc|ccccc}
\hline\hline
molecule & $n_q$ & Ansatz & $n_l$ & $d$ & $n_\theta$ & $n_g$ \\
\hline
H$_2$O & 6 & $\ry$-linear & 8 & 50 & 54 & (54,40) \\
       &   & $\ry$-full & 7 & 114 & 48 & (48,112) \\
       &   & cascade & 4 & 50 & 54 & (54,40) \\
\hline
BeH$_2$ & 7 & $\ry$-linear & 10 & 72 & 77 & (77,60) \\
 &  & $\ry$-full & 10 & 222 & 77 & (77,220) \\
 &  & cascade & 6 & 86 & 91 & (91,72) \\
\hline
HF & 6 & $\ry$-linear & 5 & 32 & 36 & (36,25) \\
 &  & $\ry$-full & 3 & 50 & 24 & (24,48) \\
 &  & cascade & 3 & 38 & 42 & (42,30) \\
\hline
BH & 6 & $\ry$-linear & 8 & 50 & 54 & (54,40) \\
 &  & $\ry$-full & 8 & 130 & 54 & (54,128) \\
 &  & cascade & 4 & 50 & 54 & (54,40) \\
\hline
LiH & 4 & $\ry$-linear & 3 & 14 & 16 & (16,9) \\
 &  & $\ry$-full & 3 & 23 & 16 & (16,21) \\
 &  & cascade & 2 & 18 & 20 & (20,12) \\
\hline\hline
\end{tabular}

$ $

\vspace{0.5cm}

$ $

\begin{tabular}{cc|ccccc}
\hline\hline
molecule & $n_q$ & Ansatz & $n_l$ & $d$ & $n_\theta$ & $n_g$  \\
\hline
H$_2$O & 6 & $\ry$-linear & 5 & 32 & 36 & (36,25) \\
 &  & $\ry$-full & 5 & 82 & 36 & (36,80) \\
 &  & cascade & 3 & 38 & 42 & (42,30) \\
\hline
BeH$_2$ & 6 & $\ry$-linear & 5 & 32 & 36 & (36,25) \\
 &  & $\ry$-full & 5 & 82 & 36 & (36,80) \\
 &  & cascade & 3 & 38 & 42 & (42,30) \\
\hline
HF & 3 & $\ry$-linear & 5 & 17 & 18 & (18,10) \\
 &  & $\ry$-full & 5 & 22 & 18 & (18,20) \\
 &  & cascade & 3 & 20 & 21 & (21,12) \\
\hline
BH & 5 & $\ry$-linear & 5 & 27 & 30 & (30,20) \\
 &  & $\ry$-full & 5 & 57 & 30 & (30,55) \\
 &  & cascade & 3 & 32 & 35 & (35,24) \\
\hline
LiH & 3 & $\ry$-linear & 5 & 17 & 18 & (18,10) \\
 &  & $\ry$-full & 5 & 22 & 18 & (18,20) \\
 &  & cascade & 3 & 20 & 21 & (21,12) \\
\hline\hline
\end{tabular}
\caption{Computational cost of second-quantization (top) and first-quantization (bottom, with padding scheme) simulations carried out in this work.
$n_q$ and $n_l$  indicate the number of qubits and the highest studied number of layers.
$n_{\bgreek{\theta}}$, $n_g$, and $d$ denote numbers of parameters, gates (single- and two-qubit), and circuit depth.}
\label{table:second_qtz_cost}
\end{table}

\begin{figure}[h!]
\includegraphics[width=0.65\textwidth]{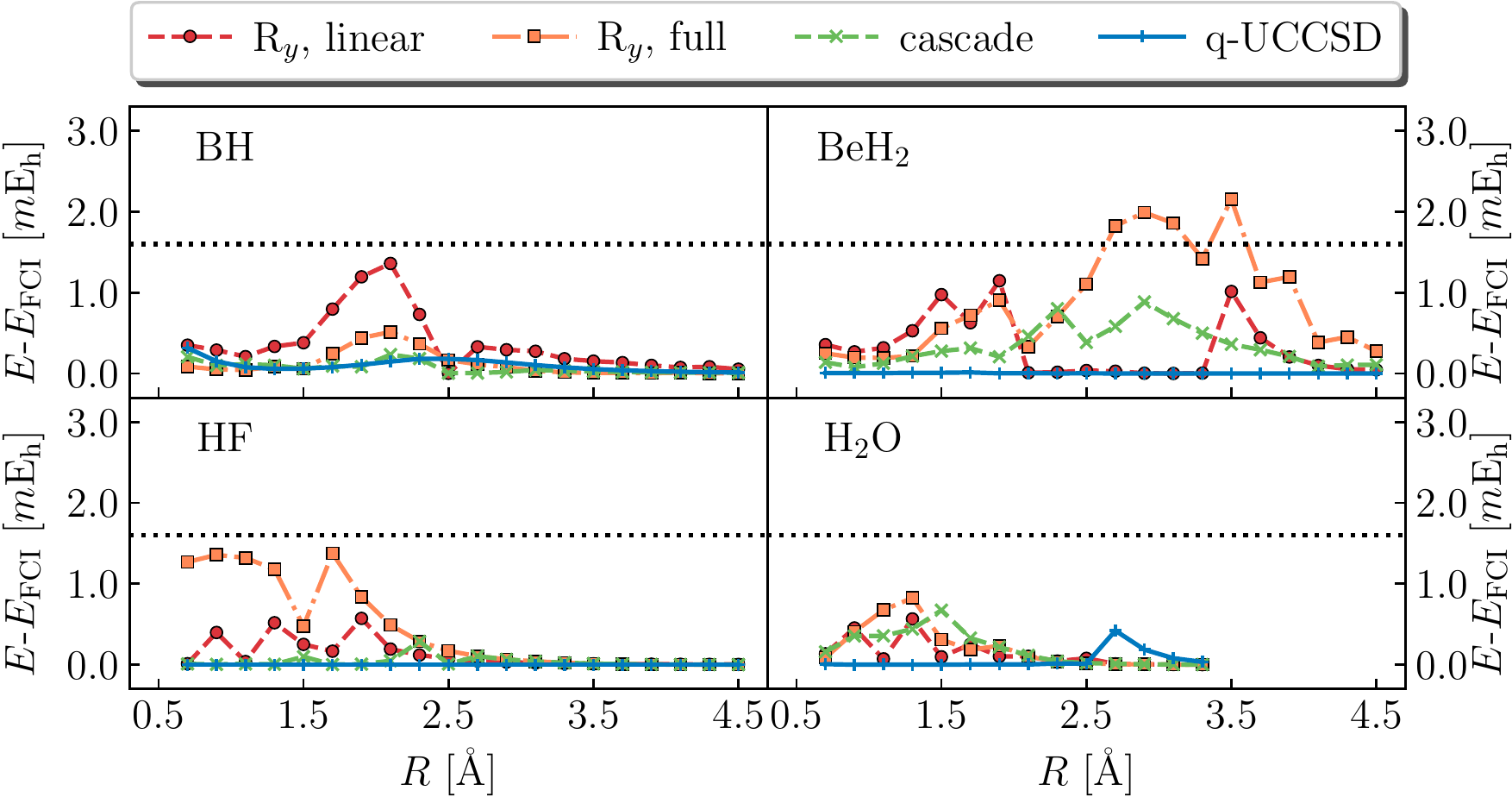}
\caption{Deviation between computed and exact total energy for the BH, HF, BeH$_2$, and H$_2$O molecules at the STO-6G level 
(counterclockwise) using the R$_y$ (with linear- and full-connectivity), cascade, and q-UCCSD Ans\"{a}tze 
(red circles, orange squares, green crosses, and blue markers) with the highest computed number of layers. The dotted black line refers to a difference with the FCI energy of 1 kcal/mol.
}
\label{figure:second_comparison}
\end{figure} 

Agreement with FCI within 1.6 milliHartree (1 kcal/mol) and smoother potential energy curves 
are only achieved, for second-quantization simulations, by increasing $n_l$ to the values listed in Table \ref{table:second_qtz_cost}.
The energies of such Ans\"{a}tze are shown in Figure \ref{figure:second_comparison}.
As seen, 1 kcal/mol was reached for all molecules except for BeH$_2$
with $R_y$ Ansatz and full-connectivity, where even $n_l=10$ leads to energies more than 1.6 milliHartree above FCI. To achieve kcal/mol accuracy, q-UCCSD calculations required $n_l=1$ layers for  BH, HF and $n_l=2$ layers for BeH$_2$, H$_2$O. This likely arises in the latter cases since two covalent bonds are broken simultaneously, and a second repetition of the q-UCCSD Ansatz 
includes the effect of disconnected quadruple excitations.

The computational cost of the circuits implementing the hardware-efficient Ans\"{a}tze in Figure \ref{figure:second_comparison} is listed in  Table \ref{table:second_qtz_cost}, along with first-quantization simulations in Figure \ref{figure:first_quantization_results} and Section V of the SI. The computational cost is estimated considering: 
number of qubits $n_q$, number of layers $n_l$,  
 number of gates $n_g$ both (1 qubit, 2 qubits),
 depth $d$ of the variational circuit, 
 number of Pauli operators in the Hamiltonian $n_p$, 
and  number of variational parameters $n_{\bgreek{\theta}}$.
{Note that $n_g$ is the number of gates in the circuit diagram.
Many topologies of quantum architectures restrict interactions to adjacent qubits, which in turn may increase the latency of quantum circuits compiled to these architectures. More practically, a network of SWAP gates has to be introduced in the circuit to allow entangling gates between non-adjacent qubits. The nature of such a SWAP network depends on the device connectivity, the compiler used, and 
the circuit structure (for example, linear-$\ry$ and cascade benefit from linear qubit connectivity whereas full-$\ry$ benefits from all-to-all connectivity).
We have chosen to base the calculation of the computational cost on the circuit, which is agnostic to any assumptions about hardware connectivity.

An important observation is that the number of variational parameters is comparable with the dimension $2^{n_q}$ of the $n_q$-qubit Hilbert space. This is an indication that hardware-efficient Ans\"{a}tze need to be highly expressive (which is achieved by covering a significant portion of the Hilbert space) so that a close approximation of the desired solution can be accessed. On the other hand, the expressivity of these Ans\"{a}tze comes at the cost of flatter cost function landscapes and therefore more difficult parameter optimization \cite{holmes2022connecting}.

We remark that the circuits required to reach 1.6 milliHartree accuracy have depth up to 114 layers of gates, which makes them challenging to implement on contemporary hardware given the high fidelity required by electronic structure.
}

\section*{Data availability}
\label{sec:github}

\begin{figure*}[h!]
\includegraphics[width=0.8\textwidth]{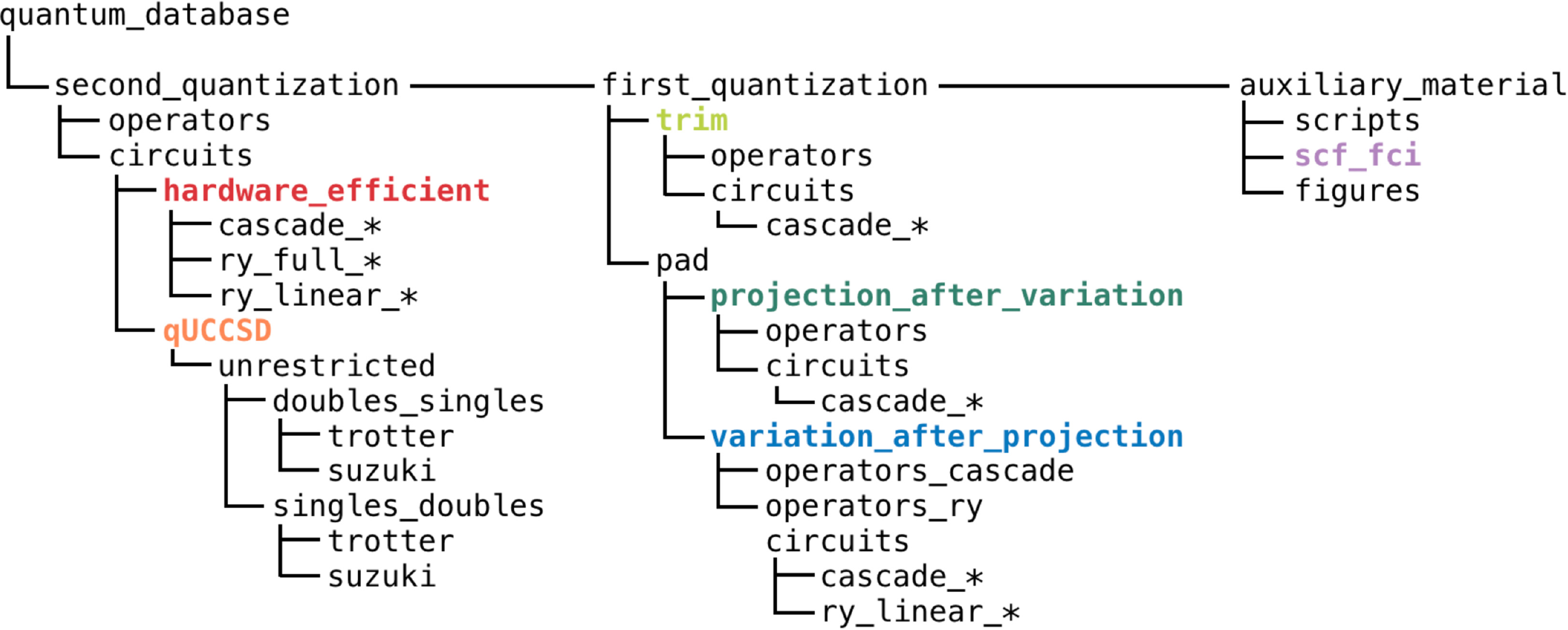}
\caption{
Organization of the repository accompanying this study.
Data under $\mathsf{hardware\_efficient}$ (red),
$\mathsf{qUCCSD}$ (orange),
$\mathsf{trim}$ (light green),
$\mathsf{projection\_after\_variation}$ (dark green),
$\mathsf{variation\_after\_projection}$ (blue)
and
$\mathsf{scf\_fci}$ (purple)
were used to generate the figures shown in the main text and the SI.
}
\label{figure:repository}
\end{figure*}

We outline the schema of a repository that stores the results to provide value to other researchers, and which can be extended in future studies to illustrate the properties of other hardware Ans\"{a}tze.

The schema is illustrated in Figure~\ref{figure:repository}.   
The results in this study, organized with respect to this schema, can be publicly accessed on GitHub at \cite{github}. Details of the format of the data within the folders can be found in Section I of the SI. 

\section{Conclusions}
\label{sec:conclusions}

In this work, we discussed some potential pitfalls connected with the use of variational quantum computing Ans\"{a}tze in electronic structure. 
{We considered Ans\"{a}tze routinely used for simulations on contemporary hardware, and studied their behavior using noiseless simulations, i.e. assuming a perfect device not affected by decoherence.}

Along with the well-known issue of challenging parameter optimization, we observed that hardware-efficient Ans\"{a}tze may break Hamiltonian symmetries and yield potential energy curves that are non-differentiable. A considerable number of layers may also be required to give results within 1 kcal/mol of FCI, thus making these calculations considerably more expensive and thus less suitable for near-term quantum hardware.

We observed that these phenomena are not perfectly correlated with each other: the analysis  of Trotterized q-UCCSD data indicates that symmetry breaking may occur in absence of challenging parameter optimization or non-differentiable potential energy curves; the analysis of first-quantization data, that challenging parameter optimization and non-differentiable potential energy curves may be observed in absence of spin symmetry breaking.

We believe these issues are important, their analysis constitutes an important criterion to design novel and refine existing hardware-efficient quantum computing algorithms, and their mitigation is an important research activity at the interface between ES and quantum computing.

While this study offers a concrete and numerical assessment of 
computational cost and algorithmic performance, it should only
be regarded as the starting point of a broader investigation.
{Extensions could include investigation of other 
variational Ans\"{a}tze \cite{parrish2019quantum,anselmetti2021local,ryabinkin2018qubit}, 
and algorithms other than VQE \cite{mcclean2017hybrid,parrish2019quantum,huggins2020non,stair2020multireference,cohn2021quantum}.
The study of molecular systems could be expanded to open-shells  (e.g. radicals), non-minimal bases (e.g. Pople \cite{ditchfield1971self} and/or correlation-consistent \cite{dunning1989gaussian} bases), and 
different properties (e.g. dissociation energy curves and spin for excited states).
Finally, from a quantum computing perspective, the evaluation of molecular observables (e.g. ground state dissociation energies and spin) could be studied using classical 
software to emulate decoherence and actual quantum devices to assess the improvements brought by 
error mitigation techniques}. 

The present study is accompanied by a repository
to integrate possible future extensions. We believe the present study will 
be a useful contribution to the understanding and development 
of quantum computing algorithms for electronic structure problems.

\section*{Supporting Information}

Contains details of the repository, information about classical preprocessing, implementation of the qUCCSD Ansatz, and additional first- and second-quantization simulations.

\section*{Acknowledgments}

MM and JER acknowledge the IBM Research Cognitive Computing Cluster service for providing resources that have contributed to the research results reported in this paper.  RD and TDC were supported by the U.S. National Science Foundation through grant CHE-2154753.

\bibliography{main}

%apsrev4-2.bst 2019-01-14 (MD) hand-edited version of apsrev4-1.bst
%Control: key (0)
%Control: author (8) initials jnrlst
%Control: editor formatted (1) identically to author
%Control: production of article title (0) allowed
%Control: page (0) single
%Control: year (1) truncated
%Control: production of eprint (0) enabled
\begin{thebibliography}{64}%
\makeatletter
\providecommand \@ifxundefined [1]{%
 \@ifx{#1\undefined}
}%
\providecommand \@ifnum [1]{%
 \ifnum #1\expandafter \@firstoftwo
 \else \expandafter \@secondoftwo
 \fi
}%
\providecommand \@ifx [1]{%
 \ifx #1\expandafter \@firstoftwo
 \else \expandafter \@secondoftwo
 \fi
}%
\providecommand \natexlab [1]{#1}%
\providecommand \enquote  [1]{``#1''}%
\providecommand \bibnamefont  [1]{#1}%
\providecommand \bibfnamefont [1]{#1}%
\providecommand \citenamefont [1]{#1}%
\providecommand \href@noop [0]{\@secondoftwo}%
\providecommand \href [0]{\begingroup \@sanitize@url \@href}%
\providecommand \@href[1]{\@@startlink{#1}\@@href}%
\providecommand \@@href[1]{\endgroup#1\@@endlink}%
\providecommand \@sanitize@url [0]{\catcode `\\12\catcode `\$12\catcode
  `\&12\catcode `\#12\catcode `\^12\catcode `\_12\catcode `\%12\relax}%
\providecommand \@@startlink[1]{}%
\providecommand \@@endlink[0]{}%
\providecommand \url  [0]{\begingroup\@sanitize@url \@url }%
\providecommand \@url [1]{\endgroup\@href {#1}{\urlprefix }}%
\providecommand \urlprefix  [0]{URL }%
\providecommand \Eprint [0]{\href }%
\providecommand \doibase [0]{https://doi.org/}%
\providecommand \selectlanguage [0]{\@gobble}%
\providecommand \bibinfo  [0]{\@secondoftwo}%
\providecommand \bibfield  [0]{\@secondoftwo}%
\providecommand \translation [1]{[#1]}%
\providecommand \BibitemOpen [0]{}%
\providecommand \bibitemStop [0]{}%
\providecommand \bibitemNoStop [0]{.\EOS\space}%
\providecommand \EOS [0]{\spacefactor3000\relax}%
\providecommand \BibitemShut  [1]{\csname bibitem#1\endcsname}%
\let\auto@bib@innerbib\@empty
%</preamble>
\bibitem [{\citenamefont {Georgescu}\ \emph {et~al.}(2014)\citenamefont
  {Georgescu}, \citenamefont {Ashhab},\ and\ \citenamefont
  {Nori}}]{georgescu2014quantum}%
  \BibitemOpen
  \bibfield  {author} {\bibinfo {author} {\bibfnamefont {I.~M.}\ \bibnamefont
  {Georgescu}}, \bibinfo {author} {\bibfnamefont {S.}~\bibnamefont {Ashhab}},\
  and\ \bibinfo {author} {\bibfnamefont {F.}~\bibnamefont {Nori}},\ }\bibfield
  {title} {\bibinfo {title} {Quantum simulation},\ }\href
  {https://journals.aps.org/rmp/abstract/10.1103/RevModPhys.86.153} {\bibfield
  {journal} {\bibinfo  {journal} {Rev. Mod. Phys}\ }\textbf {\bibinfo {volume}
  {86}},\ \bibinfo {pages} {153} (\bibinfo {year} {2014})}\BibitemShut
  {NoStop}%
\bibitem [{\citenamefont {Cao}\ \emph {et~al.}(2019)\citenamefont {Cao},
  \citenamefont {Romero}, \citenamefont {Olson}, \citenamefont {Degroote},
  \citenamefont {Johnson}, \citenamefont {Kieferov{\'a}}, \citenamefont
  {Kivlichan}, \citenamefont {Menke}, \citenamefont {Peropadre}, \citenamefont
  {Sawaya} \emph {et~al.}}]{cao2019quantum}%
  \BibitemOpen
  \bibfield  {author} {\bibinfo {author} {\bibfnamefont {Y.}~\bibnamefont
  {Cao}}, \bibinfo {author} {\bibfnamefont {J.}~\bibnamefont {Romero}},
  \bibinfo {author} {\bibfnamefont {J.~P.}\ \bibnamefont {Olson}}, \bibinfo
  {author} {\bibfnamefont {M.}~\bibnamefont {Degroote}}, \bibinfo {author}
  {\bibfnamefont {P.~D.}\ \bibnamefont {Johnson}}, \bibinfo {author}
  {\bibfnamefont {M.}~\bibnamefont {Kieferov{\'a}}}, \bibinfo {author}
  {\bibfnamefont {I.~D.}\ \bibnamefont {Kivlichan}}, \bibinfo {author}
  {\bibfnamefont {T.}~\bibnamefont {Menke}}, \bibinfo {author} {\bibfnamefont
  {B.}~\bibnamefont {Peropadre}}, \bibinfo {author} {\bibfnamefont {N.~P.}\
  \bibnamefont {Sawaya}}, \emph {et~al.},\ }\bibfield  {title} {\bibinfo
  {title} {Quantum chemistry in the age of quantum computing},\ }\href
  {https://pubs.acs.org/doi/10.1021/acs.chemrev.8b00803} {\bibfield  {journal}
  {\bibinfo  {journal} {Chem. Rev}\ }\textbf {\bibinfo {volume} {119}},\
  \bibinfo {pages} {10856} (\bibinfo {year} {2019})}\BibitemShut {NoStop}%
\bibitem [{\citenamefont {Bauer}\ \emph {et~al.}(2020)\citenamefont {Bauer},
  \citenamefont {Bravyi}, \citenamefont {Motta},\ and\ \citenamefont
  {Kin-Lic~Chan}}]{bauer2020quantum}%
  \BibitemOpen
  \bibfield  {author} {\bibinfo {author} {\bibfnamefont {B.}~\bibnamefont
  {Bauer}}, \bibinfo {author} {\bibfnamefont {S.}~\bibnamefont {Bravyi}},
  \bibinfo {author} {\bibfnamefont {M.}~\bibnamefont {Motta}},\ and\ \bibinfo
  {author} {\bibfnamefont {G.}~\bibnamefont {Kin-Lic~Chan}},\ }\bibfield
  {title} {\bibinfo {title} {Quantum algorithms for quantum chemistry and
  quantum materials science},\ }\href
  {https://pubs.acs.org/doi/10.1021/acs.chemrev.9b00829} {\bibfield  {journal}
  {\bibinfo  {journal} {Chem. Rev}\ }\textbf {\bibinfo {volume} {120}},\
  \bibinfo {pages} {12685} (\bibinfo {year} {2020})}\BibitemShut {NoStop}%
\bibitem [{\citenamefont {McArdle}\ \emph {et~al.}(2020)\citenamefont
  {McArdle}, \citenamefont {Endo}, \citenamefont {Aspuru-Guzik}, \citenamefont
  {Benjamin},\ and\ \citenamefont {Yuan}}]{mcardle2020quantum}%
  \BibitemOpen
  \bibfield  {author} {\bibinfo {author} {\bibfnamefont {S.}~\bibnamefont
  {McArdle}}, \bibinfo {author} {\bibfnamefont {S.}~\bibnamefont {Endo}},
  \bibinfo {author} {\bibfnamefont {A.}~\bibnamefont {Aspuru-Guzik}}, \bibinfo
  {author} {\bibfnamefont {S.~C.}\ \bibnamefont {Benjamin}},\ and\ \bibinfo
  {author} {\bibfnamefont {X.}~\bibnamefont {Yuan}},\ }\bibfield  {title}
  {\bibinfo {title} {Quantum computational chemistry},\ }\href
  {https://doi.org/10.1103/RevModPhys.92.015003} {\bibfield  {journal}
  {\bibinfo  {journal} {Rev. Mod. Phys}\ }\textbf {\bibinfo {volume} {92}},\
  \bibinfo {pages} {015003} (\bibinfo {year} {2020})}\BibitemShut {NoStop}%
\bibitem [{\citenamefont {Motta}\ and\ \citenamefont
  {Rice}(2021)}]{motta2021emerging}%
  \BibitemOpen
  \bibfield  {author} {\bibinfo {author} {\bibfnamefont {M.}~\bibnamefont
  {Motta}}\ and\ \bibinfo {author} {\bibfnamefont {J.~E.}\ \bibnamefont
  {Rice}},\ }\bibfield  {title} {\bibinfo {title} {Emerging quantum computing
  algorithms for quantum chemistry},\ }\href
  {https://wires.onlinelibrary.wiley.com/doi/abs/10.1002/wcms.1580} {\bibfield
  {journal} {\bibinfo  {journal} {WIREs Comput. Mol. Sci}\ }\textbf {\bibinfo
  {volume} {12}},\ \bibinfo {pages} {e1580} (\bibinfo {year}
  {2021})}\BibitemShut {NoStop}%
\bibitem [{\citenamefont {Cerezo}\ \emph {et~al.}(2022)\citenamefont {Cerezo},
  \citenamefont {Arrasmith}, \citenamefont {Babbush}, \citenamefont {Benjamin},
  \citenamefont {Endo}, \citenamefont {Fujii}, \citenamefont {McClean},
  \citenamefont {Mitarai}, \citenamefont {Yuan}, \citenamefont {Cincio} \emph
  {et~al.}}]{cerezo2020variational}%
  \BibitemOpen
  \bibfield  {author} {\bibinfo {author} {\bibfnamefont {M.}~\bibnamefont
  {Cerezo}}, \bibinfo {author} {\bibfnamefont {A.}~\bibnamefont {Arrasmith}},
  \bibinfo {author} {\bibfnamefont {R.}~\bibnamefont {Babbush}}, \bibinfo
  {author} {\bibfnamefont {S.~C.}\ \bibnamefont {Benjamin}}, \bibinfo {author}
  {\bibfnamefont {S.}~\bibnamefont {Endo}}, \bibinfo {author} {\bibfnamefont
  {K.}~\bibnamefont {Fujii}}, \bibinfo {author} {\bibfnamefont {J.~R.}\
  \bibnamefont {McClean}}, \bibinfo {author} {\bibfnamefont {K.}~\bibnamefont
  {Mitarai}}, \bibinfo {author} {\bibfnamefont {X.}~\bibnamefont {Yuan}},
  \bibinfo {author} {\bibfnamefont {L.}~\bibnamefont {Cincio}}, \emph
  {et~al.},\ }\bibfield  {title} {\bibinfo {title} {Variational quantum
  algorithms},\ }\href {https://www.nature.com/articles/s42254-021-00348-9}
  {\bibfield  {journal} {\bibinfo  {journal} {Nat. Rev. Phys}\ }\textbf
  {\bibinfo {volume} {3}},\ \bibinfo {pages} {625–644} (\bibinfo {year}
  {2022})}\BibitemShut {NoStop}%
\bibitem [{\citenamefont {O’Malley}\ \emph {et~al.}(2016)\citenamefont
  {O’Malley}, \citenamefont {Babbush}, \citenamefont {Kivlichan},
  \citenamefont {Romero}, \citenamefont {McClean}, \citenamefont {Barends},
  \citenamefont {Kelly}, \citenamefont {Roushan}, \citenamefont {Tranter},
  \citenamefont {Ding} \emph {et~al.}}]{o2016scalable}%
  \BibitemOpen
  \bibfield  {author} {\bibinfo {author} {\bibfnamefont {P.~J.}\ \bibnamefont
  {O’Malley}}, \bibinfo {author} {\bibfnamefont {R.}~\bibnamefont {Babbush}},
  \bibinfo {author} {\bibfnamefont {I.~D.}\ \bibnamefont {Kivlichan}}, \bibinfo
  {author} {\bibfnamefont {J.}~\bibnamefont {Romero}}, \bibinfo {author}
  {\bibfnamefont {J.~R.}\ \bibnamefont {McClean}}, \bibinfo {author}
  {\bibfnamefont {R.}~\bibnamefont {Barends}}, \bibinfo {author} {\bibfnamefont
  {J.}~\bibnamefont {Kelly}}, \bibinfo {author} {\bibfnamefont
  {P.}~\bibnamefont {Roushan}}, \bibinfo {author} {\bibfnamefont
  {A.}~\bibnamefont {Tranter}}, \bibinfo {author} {\bibfnamefont
  {N.}~\bibnamefont {Ding}}, \emph {et~al.},\ }\bibfield  {title} {\bibinfo
  {title} {Scalable quantum simulation of molecular energies},\ }\href
  {https://journals.aps.org/prx/abstract/10.1103/PhysRevX.6.031007} {\bibfield
  {journal} {\bibinfo  {journal} {Phys. Rev. X}\ }\textbf {\bibinfo {volume}
  {6}},\ \bibinfo {pages} {031007} (\bibinfo {year} {2016})}\BibitemShut
  {NoStop}%
\bibitem [{\citenamefont {Kandala}\ \emph {et~al.}(2017)\citenamefont
  {Kandala}, \citenamefont {Mezzacapo}, \citenamefont {Temme}, \citenamefont
  {Takita}, \citenamefont {Brink}, \citenamefont {Chow},\ and\ \citenamefont
  {Gambetta}}]{kandala2017hardware}%
  \BibitemOpen
  \bibfield  {author} {\bibinfo {author} {\bibfnamefont {A.}~\bibnamefont
  {Kandala}}, \bibinfo {author} {\bibfnamefont {A.}~\bibnamefont {Mezzacapo}},
  \bibinfo {author} {\bibfnamefont {K.}~\bibnamefont {Temme}}, \bibinfo
  {author} {\bibfnamefont {M.}~\bibnamefont {Takita}}, \bibinfo {author}
  {\bibfnamefont {M.}~\bibnamefont {Brink}}, \bibinfo {author} {\bibfnamefont
  {J.~M.}\ \bibnamefont {Chow}},\ and\ \bibinfo {author} {\bibfnamefont
  {J.~M.}\ \bibnamefont {Gambetta}},\ }\bibfield  {title} {\bibinfo {title}
  {Hardware-efficient variational quantum eigensolver for small molecules and
  quantum magnets},\ }\href {https://doi.org/10.1038/nature23879} {\bibfield
  {journal} {\bibinfo  {journal} {Nature}\ }\textbf {\bibinfo {volume} {549}},\
  \bibinfo {pages} {242} (\bibinfo {year} {2017})}\BibitemShut {NoStop}%
\bibitem [{\citenamefont {Arute}\ \emph {et~al.}(2020)\citenamefont {Arute}
  \emph {et~al.}}]{google2020hartree}%
  \BibitemOpen
  \bibfield  {author} {\bibinfo {author} {\bibfnamefont {F.}~\bibnamefont
  {Arute}} \emph {et~al.},\ }\bibfield  {title} {\bibinfo {title}
  {Hartree-{F}ock on a superconducting qubit quantum computer},\ }\href
  {https://science.sciencemag.org/content/369/6507/1084} {\bibfield  {journal}
  {\bibinfo  {journal} {Science}\ }\textbf {\bibinfo {volume} {369}},\ \bibinfo
  {pages} {1084} (\bibinfo {year} {2020})}\BibitemShut {NoStop}%
\bibitem [{\citenamefont {Rice}\ \emph {et~al.}(2021)\citenamefont {Rice},
  \citenamefont {Gujarati}, \citenamefont {Motta}, \citenamefont {Takeshita},
  \citenamefont {Lee}, \citenamefont {Latone},\ and\ \citenamefont
  {Garcia}}]{rice2021quantum}%
  \BibitemOpen
  \bibfield  {author} {\bibinfo {author} {\bibfnamefont {J.~E.}\ \bibnamefont
  {Rice}}, \bibinfo {author} {\bibfnamefont {T.~P.}\ \bibnamefont {Gujarati}},
  \bibinfo {author} {\bibfnamefont {M.}~\bibnamefont {Motta}}, \bibinfo
  {author} {\bibfnamefont {T.~Y.}\ \bibnamefont {Takeshita}}, \bibinfo {author}
  {\bibfnamefont {E.}~\bibnamefont {Lee}}, \bibinfo {author} {\bibfnamefont
  {J.~A.}\ \bibnamefont {Latone}},\ and\ \bibinfo {author} {\bibfnamefont
  {J.~M.}\ \bibnamefont {Garcia}},\ }\bibfield  {title} {\bibinfo {title}
  {Quantum computation of dominant products in lithium--sulfur batteries},\
  }\href {https://aip.scitation.org/doi/10.1063/5.0044068} {\bibfield
  {journal} {\bibinfo  {journal} {J. Chem. Phys}\ }\textbf {\bibinfo {volume}
  {154}},\ \bibinfo {pages} {134115} (\bibinfo {year} {2021})}\BibitemShut
  {NoStop}%
\bibitem [{\citenamefont {Peruzzo}\ \emph {et~al.}(2014)\citenamefont
  {Peruzzo}, \citenamefont {McClean}, \citenamefont {Shadbolt}, \citenamefont
  {Yung}, \citenamefont {Zhou}, \citenamefont {Love}, \citenamefont
  {Aspuru-Guzik},\ and\ \citenamefont {O'Brien}}]{peruzzo2014variational}%
  \BibitemOpen
  \bibfield  {author} {\bibinfo {author} {\bibfnamefont {A.}~\bibnamefont
  {Peruzzo}}, \bibinfo {author} {\bibfnamefont {J.}~\bibnamefont {McClean}},
  \bibinfo {author} {\bibfnamefont {P.}~\bibnamefont {Shadbolt}}, \bibinfo
  {author} {\bibfnamefont {M.-H.}\ \bibnamefont {Yung}}, \bibinfo {author}
  {\bibfnamefont {X.-Q.}\ \bibnamefont {Zhou}}, \bibinfo {author}
  {\bibfnamefont {P.~J.}\ \bibnamefont {Love}}, \bibinfo {author}
  {\bibfnamefont {A.}~\bibnamefont {Aspuru-Guzik}},\ and\ \bibinfo {author}
  {\bibfnamefont {J.~L.}\ \bibnamefont {O'Brien}},\ }\bibfield  {title}
  {\bibinfo {title} {A variational eigenvalue solver on a photonic quantum
  processor},\ }\href {https://doi.org/10.1038/ncomms5213} {\bibfield
  {journal} {\bibinfo  {journal} {Nat. Commun}\ }\textbf {\bibinfo {volume}
  {5}},\ \bibinfo {pages} {4213} (\bibinfo {year} {2014})}\BibitemShut
  {NoStop}%
\bibitem [{\citenamefont {McClean}\ \emph {et~al.}(2018)\citenamefont
  {McClean}, \citenamefont {Boixo}, \citenamefont {Smelyanskiy}, \citenamefont
  {Babbush},\ and\ \citenamefont {Neven}}]{mcclean2018barren}%
  \BibitemOpen
  \bibfield  {author} {\bibinfo {author} {\bibfnamefont {J.~R.}\ \bibnamefont
  {McClean}}, \bibinfo {author} {\bibfnamefont {S.}~\bibnamefont {Boixo}},
  \bibinfo {author} {\bibfnamefont {V.~N.}\ \bibnamefont {Smelyanskiy}},
  \bibinfo {author} {\bibfnamefont {R.}~\bibnamefont {Babbush}},\ and\ \bibinfo
  {author} {\bibfnamefont {H.}~\bibnamefont {Neven}},\ }\bibfield  {title}
  {\bibinfo {title} {Barren plateaus in quantum neural network training
  landscapes},\ }\href {https://www.nature.com/articles/s41467-018-07090-4}
  {\bibfield  {journal} {\bibinfo  {journal} {Nat. Commun}\ }\textbf {\bibinfo
  {volume} {9}},\ \bibinfo {pages} {4812} (\bibinfo {year} {2018})}\BibitemShut
  {NoStop}%
\bibitem [{\citenamefont {Gomes}\ \emph {et~al.}(2020)\citenamefont {Gomes},
  \citenamefont {Zhang}, \citenamefont {Berthusen}, \citenamefont {Wang},
  \citenamefont {Ho}, \citenamefont {Orth},\ and\ \citenamefont
  {Yao}}]{gomes2020efficient}%
  \BibitemOpen
  \bibfield  {author} {\bibinfo {author} {\bibfnamefont {N.}~\bibnamefont
  {Gomes}}, \bibinfo {author} {\bibfnamefont {F.}~\bibnamefont {Zhang}},
  \bibinfo {author} {\bibfnamefont {N.~F.}\ \bibnamefont {Berthusen}}, \bibinfo
  {author} {\bibfnamefont {C.-Z.}\ \bibnamefont {Wang}}, \bibinfo {author}
  {\bibfnamefont {K.-M.}\ \bibnamefont {Ho}}, \bibinfo {author} {\bibfnamefont
  {P.~P.}\ \bibnamefont {Orth}},\ and\ \bibinfo {author} {\bibfnamefont
  {Y.}~\bibnamefont {Yao}},\ }\bibfield  {title} {\bibinfo {title} {Efficient
  step-merged quantum imaginary time evolution algorithm for quantum
  chemistry},\ }\href {https://pubs.acs.org/doi/abs/10.1021/acs.jctc.0c00666}
  {\bibfield  {journal} {\bibinfo  {journal} {J. Chem. Theory Comput}\ }\textbf
  {\bibinfo {volume} {16}},\ \bibinfo {pages} {6256} (\bibinfo {year}
  {2020})}\BibitemShut {NoStop}%
\bibitem [{\citenamefont {Aharonov}\ and\ \citenamefont
  {Ta-Shma}(2003)}]{aharonov2003adiabatic}%
  \BibitemOpen
  \bibfield  {author} {\bibinfo {author} {\bibfnamefont {D.}~\bibnamefont
  {Aharonov}}\ and\ \bibinfo {author} {\bibfnamefont {A.}~\bibnamefont
  {Ta-Shma}},\ }\bibfield  {title} {\bibinfo {title} {Adiabatic quantum state
  generation and statistical zero knowledge},\ }in\ \href
  {https://dl.acm.org/doi/10.1145/780542.780546} {\emph {\bibinfo {booktitle}
  {Proc. ACM}}}\ (\bibinfo {year} {2003})\ pp.\ \bibinfo {pages}
  {20--29}\BibitemShut {NoStop}%
\bibitem [{\citenamefont {Kitaev}(1995)}]{kitaev1995quantum}%
  \BibitemOpen
  \bibfield  {author} {\bibinfo {author} {\bibfnamefont {A.~Y.}\ \bibnamefont
  {Kitaev}},\ }\bibfield  {title} {\bibinfo {title} {Quantum measurements and
  the {Abelian} stabilizer problem},\ }\href
  {https://arxiv.org/abs/quant-ph/9511026} {\bibfield  {journal} {\bibinfo
  {journal} {arXiv quant-ph/9511026}\ } (\bibinfo {year} {1995})}\BibitemShut
  {NoStop}%
\bibitem [{\citenamefont {Seeley}\ \emph {et~al.}(2012)\citenamefont {Seeley},
  \citenamefont {Richard},\ and\ \citenamefont {Love}}]{seeley2012bravyi}%
  \BibitemOpen
  \bibfield  {author} {\bibinfo {author} {\bibfnamefont {J.~T.}\ \bibnamefont
  {Seeley}}, \bibinfo {author} {\bibfnamefont {M.~J.}\ \bibnamefont
  {Richard}},\ and\ \bibinfo {author} {\bibfnamefont {P.~J.}\ \bibnamefont
  {Love}},\ }\bibfield  {title} {\bibinfo {title} {The {B}ravyi-{K}itaev
  transformation for quantum computation of electronic structure},\ }\href
  {https://aip.scitation.org/doi/10.1063/1.4768229} {\bibfield  {journal}
  {\bibinfo  {journal} {J. Chem. Phys}\ }\textbf {\bibinfo {volume} {137}},\
  \bibinfo {pages} {224109} (\bibinfo {year} {2012})}\BibitemShut {NoStop}%
\bibitem [{\citenamefont {Bravyi}\ and\ \citenamefont
  {Kitaev}(2002)}]{bravyi2002fermionic}%
  \BibitemOpen
  \bibfield  {author} {\bibinfo {author} {\bibfnamefont {S.~B.}\ \bibnamefont
  {Bravyi}}\ and\ \bibinfo {author} {\bibfnamefont {A.~Y.}\ \bibnamefont
  {Kitaev}},\ }\bibfield  {title} {\bibinfo {title} {Fermionic quantum
  computation},\ }\href
  {https://www.sciencedirect.com/science/article/abs/pii/S0003491602962548}
  {\bibfield  {journal} {\bibinfo  {journal} {Ann. Phys}\ }\textbf {\bibinfo
  {volume} {298}},\ \bibinfo {pages} {210} (\bibinfo {year}
  {2002})}\BibitemShut {NoStop}%
\bibitem [{\citenamefont {Bravyi}\ \emph {et~al.}(2017)\citenamefont {Bravyi},
  \citenamefont {Gambetta}, \citenamefont {Mezzacapo},\ and\ \citenamefont
  {Temme}}]{bravyi2017tapering}%
  \BibitemOpen
  \bibfield  {author} {\bibinfo {author} {\bibfnamefont {S.}~\bibnamefont
  {Bravyi}}, \bibinfo {author} {\bibfnamefont {J.~M.}\ \bibnamefont
  {Gambetta}}, \bibinfo {author} {\bibfnamefont {A.}~\bibnamefont
  {Mezzacapo}},\ and\ \bibinfo {author} {\bibfnamefont {K.}~\bibnamefont
  {Temme}},\ }\bibfield  {title} {\bibinfo {title} {Tapering off qubits to
  simulate fermionic hamiltonians},\ }\href {https://arxiv.org/abs/1701.08213}
  {\bibfield  {journal} {\bibinfo  {journal} {arXiv:1701.08213}\ } (\bibinfo
  {year} {2017})}\BibitemShut {NoStop}%
\bibitem [{\citenamefont {Setia}\ \emph {et~al.}(2020)\citenamefont {Setia},
  \citenamefont {Chen}, \citenamefont {Rice}, \citenamefont {Mezzacapo},
  \citenamefont {Pistoia},\ and\ \citenamefont
  {Whitfield}}]{setia2019reducing}%
  \BibitemOpen
  \bibfield  {author} {\bibinfo {author} {\bibfnamefont {K.}~\bibnamefont
  {Setia}}, \bibinfo {author} {\bibfnamefont {R.}~\bibnamefont {Chen}},
  \bibinfo {author} {\bibfnamefont {J.~E.}\ \bibnamefont {Rice}}, \bibinfo
  {author} {\bibfnamefont {A.}~\bibnamefont {Mezzacapo}}, \bibinfo {author}
  {\bibfnamefont {M.}~\bibnamefont {Pistoia}},\ and\ \bibinfo {author}
  {\bibfnamefont {J.~D.}\ \bibnamefont {Whitfield}},\ }\bibfield  {title}
  {\bibinfo {title} {Reducing qubit requirements for quantum simulations using
  molecular point group symmetries},\ }\href
  {https://pubs.acs.org/doi/10.1021/acs.jctc.0c00113} {\bibfield  {journal}
  {\bibinfo  {journal} {J. Chem. Theory Comput}\ }\textbf {\bibinfo {volume}
  {16}},\ \bibinfo {pages} {6091} (\bibinfo {year} {2020})}\BibitemShut
  {NoStop}%
\bibitem [{\citenamefont {Fischer}\ and\ \citenamefont
  {Gunlycke}(2019)}]{fischer2019symmetry}%
  \BibitemOpen
  \bibfield  {author} {\bibinfo {author} {\bibfnamefont {S.~A.}\ \bibnamefont
  {Fischer}}\ and\ \bibinfo {author} {\bibfnamefont {D.}~\bibnamefont
  {Gunlycke}},\ }\bibfield  {title} {\bibinfo {title} {Symmetry configuration
  mapping for representing quantum systems on quantum computers},\ }\href
  {https://arxiv.org/abs/1907.01493} {\bibfield  {journal} {\bibinfo  {journal}
  {arXiv:1907.01493}\ } (\bibinfo {year} {2019})}\BibitemShut {NoStop}%
\bibitem [{\citenamefont {Sugisaki}\ \emph {et~al.}(2018)\citenamefont
  {Sugisaki}, \citenamefont {Nakazawa}, \citenamefont {Toyota}, \citenamefont
  {Sato}, \citenamefont {Shiomi},\ and\ \citenamefont
  {Takui}}]{sugisaki2018quantum}%
  \BibitemOpen
  \bibfield  {author} {\bibinfo {author} {\bibfnamefont {K.}~\bibnamefont
  {Sugisaki}}, \bibinfo {author} {\bibfnamefont {S.}~\bibnamefont {Nakazawa}},
  \bibinfo {author} {\bibfnamefont {K.}~\bibnamefont {Toyota}}, \bibinfo
  {author} {\bibfnamefont {K.}~\bibnamefont {Sato}}, \bibinfo {author}
  {\bibfnamefont {D.}~\bibnamefont {Shiomi}},\ and\ \bibinfo {author}
  {\bibfnamefont {T.}~\bibnamefont {Takui}},\ }\bibfield  {title} {\bibinfo
  {title} {Quantum chemistry on quantum computers: a method for preparation of
  multiconfigurational wave functions on quantum computers without performing
  post-hartree--fock calculations},\ }\href
  {https://pubs.acs.org/doi/10.1021/acscentsci.8b00788} {\bibfield  {journal}
  {\bibinfo  {journal} {ACS Central Sci}\ }\textbf {\bibinfo {volume} {5}},\
  \bibinfo {pages} {167} (\bibinfo {year} {2018})}\BibitemShut {NoStop}%
\bibitem [{\citenamefont {Sugisaki}\ \emph {et~al.}(2019)\citenamefont
  {Sugisaki}, \citenamefont {Yamamoto}, \citenamefont {Nakazawa}, \citenamefont
  {Toyota}, \citenamefont {Sato}, \citenamefont {Shiomi},\ and\ \citenamefont
  {Takui}}]{sugisaki2019open}%
  \BibitemOpen
  \bibfield  {author} {\bibinfo {author} {\bibfnamefont {K.}~\bibnamefont
  {Sugisaki}}, \bibinfo {author} {\bibfnamefont {S.}~\bibnamefont {Yamamoto}},
  \bibinfo {author} {\bibfnamefont {S.}~\bibnamefont {Nakazawa}}, \bibinfo
  {author} {\bibfnamefont {K.}~\bibnamefont {Toyota}}, \bibinfo {author}
  {\bibfnamefont {K.}~\bibnamefont {Sato}}, \bibinfo {author} {\bibfnamefont
  {D.}~\bibnamefont {Shiomi}},\ and\ \bibinfo {author} {\bibfnamefont
  {T.}~\bibnamefont {Takui}},\ }\bibfield  {title} {\bibinfo {title} {Open
  shell electronic state calculations on quantum computers: A quantum circuit
  for the preparation of configuration state functions based on serber
  construction},\ }\href
  {https://linkinghub.elsevier.com/retrieve/pii/S2590141918300023} {\bibfield
  {journal} {\bibinfo  {journal} {Chem. Phys. Lett}\ }\textbf {\bibinfo
  {volume} {737}},\ \bibinfo {pages} {100002} (\bibinfo {year}
  {2019})}\BibitemShut {NoStop}%
\bibitem [{\citenamefont {Gunlycke}\ \emph {et~al.}(2020)\citenamefont
  {Gunlycke}, \citenamefont {Palenik}, \citenamefont {Emmert},\ and\
  \citenamefont {Fischer}}]{gunlycke2020efficient}%
  \BibitemOpen
  \bibfield  {author} {\bibinfo {author} {\bibfnamefont {D.}~\bibnamefont
  {Gunlycke}}, \bibinfo {author} {\bibfnamefont {M.~C.}\ \bibnamefont
  {Palenik}}, \bibinfo {author} {\bibfnamefont {A.~R.}\ \bibnamefont
  {Emmert}},\ and\ \bibinfo {author} {\bibfnamefont {S.~A.}\ \bibnamefont
  {Fischer}},\ }\bibfield  {title} {\bibinfo {title} {Efficient algorithm for
  generating pauli coordinates for an arbitrary linear operator},\ }\href
  {https://arxiv.org/abs/2011.08942} {\bibfield  {journal} {\bibinfo  {journal}
  {arXiv:2011.08942}\ } (\bibinfo {year} {2020})}\BibitemShut {NoStop}%
\bibitem [{\citenamefont {Gunlycke}(2021)}]{gunlycke2021compact}%
  \BibitemOpen
  \bibfield  {author} {\bibinfo {author} {\bibfnamefont {L.~D.}\ \bibnamefont
  {Gunlycke}},\ }\href {https://patents.google.com/patent/US20210263753A1/en}
  {\bibinfo {title} {Compact, symmetry-adapted mapping between fermionic
  systems and quantum computers}} (\bibinfo {year} {2021}),\ \bibinfo {note}
  {{U}{S} Patent App. 17/184,516}\BibitemShut {NoStop}%
\bibitem [{\citenamefont {Scuseria}\ \emph {et~al.}(2011)\citenamefont
  {Scuseria}, \citenamefont {Jim{\'e}nez-Hoyos}, \citenamefont {Henderson},
  \citenamefont {Samanta},\ and\ \citenamefont
  {Ellis}}]{scuseria2011projected}%
  \BibitemOpen
  \bibfield  {author} {\bibinfo {author} {\bibfnamefont {G.~E.}\ \bibnamefont
  {Scuseria}}, \bibinfo {author} {\bibfnamefont {C.~A.}\ \bibnamefont
  {Jim{\'e}nez-Hoyos}}, \bibinfo {author} {\bibfnamefont {T.~M.}\ \bibnamefont
  {Henderson}}, \bibinfo {author} {\bibfnamefont {K.}~\bibnamefont {Samanta}},\
  and\ \bibinfo {author} {\bibfnamefont {J.~K.}\ \bibnamefont {Ellis}},\
  }\bibfield  {title} {\bibinfo {title} {Projected quasiparticle theory for
  molecular electronic structure},\ }\href
  {https://aip.scitation.org/doi/10.1063/1.3643338} {\bibfield  {journal}
  {\bibinfo  {journal} {J. Chem. Phys}\ }\textbf {\bibinfo {volume} {135}},\
  \bibinfo {pages} {124108} (\bibinfo {year} {2011})}\BibitemShut {NoStop}%
\bibitem [{\citenamefont {Jim{\'e}nez-Hoyos}\ \emph {et~al.}(2012)\citenamefont
  {Jim{\'e}nez-Hoyos}, \citenamefont {Henderson}, \citenamefont {Tsuchimochi},\
  and\ \citenamefont {Scuseria}}]{jimenez2012projected}%
  \BibitemOpen
  \bibfield  {author} {\bibinfo {author} {\bibfnamefont {C.~A.}\ \bibnamefont
  {Jim{\'e}nez-Hoyos}}, \bibinfo {author} {\bibfnamefont {T.~M.}\ \bibnamefont
  {Henderson}}, \bibinfo {author} {\bibfnamefont {T.}~\bibnamefont
  {Tsuchimochi}},\ and\ \bibinfo {author} {\bibfnamefont {G.~E.}\ \bibnamefont
  {Scuseria}},\ }\bibfield  {title} {\bibinfo {title} {Projected
  {H}artree--{F}ock theory},\ }\href
  {https://aip.scitation.org/doi/10.1063/1.4705280} {\bibfield  {journal}
  {\bibinfo  {journal} {J. Chem. Phys}\ }\textbf {\bibinfo {volume} {136}},\
  \bibinfo {pages} {164109} (\bibinfo {year} {2012})}\BibitemShut {NoStop}%
\bibitem [{\citenamefont {McDouall}\ and\ \citenamefont
  {Schlegel}(1989)}]{mcdouall1989analytical}%
  \BibitemOpen
  \bibfield  {author} {\bibinfo {author} {\bibfnamefont {J.~J.}\ \bibnamefont
  {McDouall}}\ and\ \bibinfo {author} {\bibfnamefont {H.~B.}\ \bibnamefont
  {Schlegel}},\ }\bibfield  {title} {\bibinfo {title} {Analytical gradients for
  unrestricted hartree--fock and second order moller--plesset perturbation
  theory with single spin annihilation},\ }\href
  {https://aip.scitation.org/doi/pdf/10.1063/1.455978} {\bibfield  {journal}
  {\bibinfo  {journal} {J. Chem. Phys}\ }\textbf {\bibinfo {volume} {90}},\
  \bibinfo {pages} {2363} (\bibinfo {year} {1989})}\BibitemShut {NoStop}%
\bibitem [{\citenamefont {Hratchian}(2013)}]{hratchian2013communication}%
  \BibitemOpen
  \bibfield  {author} {\bibinfo {author} {\bibfnamefont {H.~P.}\ \bibnamefont
  {Hratchian}},\ }\bibfield  {title} {\bibinfo {title} {Communication: An
  efficient analytic gradient theory for approximate spin projection methods},\
  }\href {https://aip.scitation.org/doi/10.1063/1.4795429} {\bibfield
  {journal} {\bibinfo  {journal} {J. Chem. Phys}\ }\textbf {\bibinfo {volume}
  {138}},\ \bibinfo {pages} {101101} (\bibinfo {year} {2013})}\BibitemShut
  {NoStop}%
\bibitem [{\citenamefont {Thompson}\ and\ \citenamefont
  {Hratchian}(2015)}]{thompson2015second}%
  \BibitemOpen
  \bibfield  {author} {\bibinfo {author} {\bibfnamefont {L.~M.}\ \bibnamefont
  {Thompson}}\ and\ \bibinfo {author} {\bibfnamefont {H.~P.}\ \bibnamefont
  {Hratchian}},\ }\bibfield  {title} {\bibinfo {title} {Second derivatives for
  approximate spin projection methods},\ }\href
  {https://aip.scitation.org/doi/10.1063/1.4907269} {\bibfield  {journal}
  {\bibinfo  {journal} {J. Chem. Phys}\ }\textbf {\bibinfo {volume} {142}},\
  \bibinfo {pages} {054106} (\bibinfo {year} {2015})}\BibitemShut {NoStop}%
\bibitem [{\citenamefont {Farhi}\ \emph {et~al.}(2014)\citenamefont {Farhi},
  \citenamefont {Goldstone},\ and\ \citenamefont {Gutmann}}]{farhi2014quantum}%
  \BibitemOpen
  \bibfield  {author} {\bibinfo {author} {\bibfnamefont {E.}~\bibnamefont
  {Farhi}}, \bibinfo {author} {\bibfnamefont {J.}~\bibnamefont {Goldstone}},\
  and\ \bibinfo {author} {\bibfnamefont {S.}~\bibnamefont {Gutmann}},\
  }\bibfield  {title} {\bibinfo {title} {A quantum approximate optimization
  algorithm},\ }\href {https://arxiv.org/abs/1411.4028} {\bibfield  {journal}
  {\bibinfo  {journal} {arXiv:1411.4028}\ } (\bibinfo {year}
  {2014})}\BibitemShut {NoStop}%
\bibitem [{\citenamefont {Ryabinkin}\ \emph {et~al.}(2018)\citenamefont
  {Ryabinkin}, \citenamefont {Yen}, \citenamefont {Genin},\ and\ \citenamefont
  {Izmaylov}}]{ryabinkin2018qubit}%
  \BibitemOpen
  \bibfield  {author} {\bibinfo {author} {\bibfnamefont {I.~G.}\ \bibnamefont
  {Ryabinkin}}, \bibinfo {author} {\bibfnamefont {T.-C.}\ \bibnamefont {Yen}},
  \bibinfo {author} {\bibfnamefont {S.~N.}\ \bibnamefont {Genin}},\ and\
  \bibinfo {author} {\bibfnamefont {A.~F.}\ \bibnamefont {Izmaylov}},\
  }\bibfield  {title} {\bibinfo {title} {Qubit coupled cluster method: a
  systematic approach to quantum chemistry on a quantum computer},\ }\href
  {https://pubs.acs.org/doi/10.1021/acs.jctc.8b00932} {\bibfield  {journal}
  {\bibinfo  {journal} {J. Chem. Theory Comput}\ }\textbf {\bibinfo {volume}
  {14}},\ \bibinfo {pages} {6317} (\bibinfo {year} {2018})}\BibitemShut
  {NoStop}%
\bibitem [{\citenamefont {Bartlett}\ \emph {et~al.}(1989)\citenamefont
  {Bartlett}, \citenamefont {Kucharski},\ and\ \citenamefont
  {Noga}}]{bartlett1989alternative}%
  \BibitemOpen
  \bibfield  {author} {\bibinfo {author} {\bibfnamefont {R.~J.}\ \bibnamefont
  {Bartlett}}, \bibinfo {author} {\bibfnamefont {S.~A.}\ \bibnamefont
  {Kucharski}},\ and\ \bibinfo {author} {\bibfnamefont {J.}~\bibnamefont
  {Noga}},\ }\bibfield  {title} {\bibinfo {title} {Alternative coupled-cluster
  {A}ns{\"a}tze {II}. {T}he unitary coupled-cluster method},\ }\href
  {https://www.sciencedirect.com/science/article/abs/pii/S0009261489873725}
  {\bibfield  {journal} {\bibinfo  {journal} {Chem. Phys. Lett}\ }\textbf
  {\bibinfo {volume} {155}},\ \bibinfo {pages} {133} (\bibinfo {year}
  {1989})}\BibitemShut {NoStop}%
\bibitem [{\citenamefont {Moll}\ \emph {et~al.}(2018)\citenamefont {Moll},
  \citenamefont {Barkoutsos}, \citenamefont {Bishop}, \citenamefont {Chow},
  \citenamefont {Cross}, \citenamefont {Egger}, \citenamefont {Filipp},
  \citenamefont {Fuhrer}, \citenamefont {Gambetta}, \citenamefont {Ganzhorn},
  \citenamefont {Kandala}, \citenamefont {Mezzacapo}, \citenamefont
  {M\"{u}ller}, \citenamefont {Riess}, \citenamefont {Salis}, \citenamefont
  {Smolin}, \citenamefont {Tavernelli},\ and\ \citenamefont
  {Temme}}]{Moll_2018}%
  \BibitemOpen
  \bibfield  {author} {\bibinfo {author} {\bibfnamefont {N.}~\bibnamefont
  {Moll}}, \bibinfo {author} {\bibfnamefont {P.}~\bibnamefont {Barkoutsos}},
  \bibinfo {author} {\bibfnamefont {L.~S.}\ \bibnamefont {Bishop}}, \bibinfo
  {author} {\bibfnamefont {J.~M.}\ \bibnamefont {Chow}}, \bibinfo {author}
  {\bibfnamefont {A.}~\bibnamefont {Cross}}, \bibinfo {author} {\bibfnamefont
  {D.~J.}\ \bibnamefont {Egger}}, \bibinfo {author} {\bibfnamefont
  {S.}~\bibnamefont {Filipp}}, \bibinfo {author} {\bibfnamefont
  {A.}~\bibnamefont {Fuhrer}}, \bibinfo {author} {\bibfnamefont {J.~M.}\
  \bibnamefont {Gambetta}}, \bibinfo {author} {\bibfnamefont {M.}~\bibnamefont
  {Ganzhorn}}, \bibinfo {author} {\bibfnamefont {A.}~\bibnamefont {Kandala}},
  \bibinfo {author} {\bibfnamefont {A.}~\bibnamefont {Mezzacapo}}, \bibinfo
  {author} {\bibfnamefont {P.}~\bibnamefont {M\"{u}ller}}, \bibinfo {author}
  {\bibfnamefont {W.}~\bibnamefont {Riess}}, \bibinfo {author} {\bibfnamefont
  {G.}~\bibnamefont {Salis}}, \bibinfo {author} {\bibfnamefont
  {J.}~\bibnamefont {Smolin}}, \bibinfo {author} {\bibfnamefont
  {I.}~\bibnamefont {Tavernelli}},\ and\ \bibinfo {author} {\bibfnamefont
  {K.}~\bibnamefont {Temme}},\ }\bibfield  {title} {\bibinfo {title} {Quantum
  optimization using variational algorithms on near-term quantum devices},\
  }\href {https://doi.org/10.1088/2058-9565/aab822} {\bibfield  {journal}
  {\bibinfo  {journal} {Quantum Sci. Technol.}\ }\textbf {\bibinfo {volume}
  {3}},\ \bibinfo {pages} {030503} (\bibinfo {year} {2018})}\BibitemShut
  {NoStop}%
\bibitem [{\citenamefont {Romero}\ \emph {et~al.}(2018)\citenamefont {Romero},
  \citenamefont {Babbush}, \citenamefont {McClean}, \citenamefont {Hempel},
  \citenamefont {Love},\ and\ \citenamefont
  {Aspuru-Guzik}}]{romero2018strategies}%
  \BibitemOpen
  \bibfield  {author} {\bibinfo {author} {\bibfnamefont {J.}~\bibnamefont
  {Romero}}, \bibinfo {author} {\bibfnamefont {R.}~\bibnamefont {Babbush}},
  \bibinfo {author} {\bibfnamefont {J.~R.}\ \bibnamefont {McClean}}, \bibinfo
  {author} {\bibfnamefont {C.}~\bibnamefont {Hempel}}, \bibinfo {author}
  {\bibfnamefont {P.~J.}\ \bibnamefont {Love}},\ and\ \bibinfo {author}
  {\bibfnamefont {A.}~\bibnamefont {Aspuru-Guzik}},\ }\bibfield  {title}
  {\bibinfo {title} {Strategies for quantum computing molecular energies using
  the unitary coupled cluster {A}nsatz},\ }\href
  {https://iopscience.iop.org/article/10.1088/2058-9565/aad3e4/meta} {\bibfield
   {journal} {\bibinfo  {journal} {Quantum Sci. Technol.}\ }\textbf {\bibinfo
  {volume} {4}},\ \bibinfo {pages} {014008} (\bibinfo {year}
  {2018})}\BibitemShut {NoStop}%
\bibitem [{\citenamefont {Cooper}\ and\ \citenamefont
  {Knowles}(2010)}]{Cooper2010}%
  \BibitemOpen
  \bibfield  {author} {\bibinfo {author} {\bibfnamefont {B.}~\bibnamefont
  {Cooper}}\ and\ \bibinfo {author} {\bibfnamefont {P.~J.}\ \bibnamefont
  {Knowles}},\ }\bibfield  {title} {\bibinfo {title} {Benchmark studies of
  variational, unitary and extended coupled cluster methods},\ }\href
  {https://doi.org/10.1063/1.3520564} {\bibfield  {journal} {\bibinfo
  {journal} {J. Chem. Phys}\ }\textbf {\bibinfo {volume} {133}},\ \bibinfo
  {pages} {234102} (\bibinfo {year} {2010})}\BibitemShut {NoStop}%
\bibitem [{\citenamefont {Harsha}\ \emph {et~al.}(2018)\citenamefont {Harsha},
  \citenamefont {Shiozaki},\ and\ \citenamefont
  {Scuseria}}]{harsha2018difference}%
  \BibitemOpen
  \bibfield  {author} {\bibinfo {author} {\bibfnamefont {G.}~\bibnamefont
  {Harsha}}, \bibinfo {author} {\bibfnamefont {T.}~\bibnamefont {Shiozaki}},\
  and\ \bibinfo {author} {\bibfnamefont {G.~E.}\ \bibnamefont {Scuseria}},\
  }\bibfield  {title} {\bibinfo {title} {On the difference between variational
  and unitary coupled cluster theories},\ }\href
  {https://doi.org/10.1063/1.5011033} {\bibfield  {journal} {\bibinfo
  {journal} {J. Chem. Phys}\ }\textbf {\bibinfo {volume} {148}},\ \bibinfo
  {pages} {044107} (\bibinfo {year} {2018})}\BibitemShut {NoStop}%
\bibitem [{\citenamefont {Evangelista}\ \emph {et~al.}(2019)\citenamefont
  {Evangelista}, \citenamefont {Chan},\ and\ \citenamefont
  {Scuseria}}]{evangelista2019exact}%
  \BibitemOpen
  \bibfield  {author} {\bibinfo {author} {\bibfnamefont {F.~A.}\ \bibnamefont
  {Evangelista}}, \bibinfo {author} {\bibfnamefont {G.~K.-L.}\ \bibnamefont
  {Chan}},\ and\ \bibinfo {author} {\bibfnamefont {G.~E.}\ \bibnamefont
  {Scuseria}},\ }\bibfield  {title} {\bibinfo {title} {Exact parameterization
  of fermionic wave functions via unitary coupled cluster theory},\ }\href
  {https://aip.scitation.org/doi/10.1063/1.5133059} {\bibfield  {journal}
  {\bibinfo  {journal} {J. Chem. Phys}\ }\textbf {\bibinfo {volume} {151}},\
  \bibinfo {pages} {244112} (\bibinfo {year} {2019})}\BibitemShut {NoStop}%
\bibitem [{\citenamefont {Barkoutsos}\ \emph {et~al.}(2018)\citenamefont
  {Barkoutsos}, \citenamefont {Gonthier}, \citenamefont {Sokolov},
  \citenamefont {Moll}, \citenamefont {Salis}, \citenamefont {Fuhrer},
  \citenamefont {Ganzhorn}, \citenamefont {Egger}, \citenamefont {Troyer},
  \citenamefont {Mezzacapo}, \citenamefont {Filipp},\ and\ \citenamefont
  {Tavernelli}}]{barkoutsos2018quantum}%
  \BibitemOpen
  \bibfield  {author} {\bibinfo {author} {\bibfnamefont {P.~K.}\ \bibnamefont
  {Barkoutsos}}, \bibinfo {author} {\bibfnamefont {J.~F.}\ \bibnamefont
  {Gonthier}}, \bibinfo {author} {\bibfnamefont {I.}~\bibnamefont {Sokolov}},
  \bibinfo {author} {\bibfnamefont {N.}~\bibnamefont {Moll}}, \bibinfo {author}
  {\bibfnamefont {G.}~\bibnamefont {Salis}}, \bibinfo {author} {\bibfnamefont
  {A.}~\bibnamefont {Fuhrer}}, \bibinfo {author} {\bibfnamefont
  {M.}~\bibnamefont {Ganzhorn}}, \bibinfo {author} {\bibfnamefont {D.~J.}\
  \bibnamefont {Egger}}, \bibinfo {author} {\bibfnamefont {M.}~\bibnamefont
  {Troyer}}, \bibinfo {author} {\bibfnamefont {A.}~\bibnamefont {Mezzacapo}},
  \bibinfo {author} {\bibfnamefont {S.}~\bibnamefont {Filipp}},\ and\ \bibinfo
  {author} {\bibfnamefont {I.}~\bibnamefont {Tavernelli}},\ }\bibfield  {title}
  {\bibinfo {title} {Quantum algorithms for electronic structure calculations:
  Particle-hole hamiltonian and optimized wave-function expansions},\ }\href
  {https://doi.org/10.1103/PhysRevA.98.022322} {\bibfield  {journal} {\bibinfo
  {journal} {Phys. Rev. A}\ }\textbf {\bibinfo {volume} {98}},\ \bibinfo
  {pages} {022322} (\bibinfo {year} {2018})}\BibitemShut {NoStop}%
\bibitem [{\citenamefont {Paldus}(1977)}]{paldus1977correlation}%
  \BibitemOpen
  \bibfield  {author} {\bibinfo {author} {\bibfnamefont {J.}~\bibnamefont
  {Paldus}},\ }\bibfield  {title} {\bibinfo {title} {Correlation problems in
  atomic and molecular systems. v. spin-adapted coupled cluster many-electron
  theory},\ }\href {https://aip.scitation.org/doi/10.1063/1.434526} {\bibfield
  {journal} {\bibinfo  {journal} {J. Chem. Phys}\ }\textbf {\bibinfo {volume}
  {67}},\ \bibinfo {pages} {303} (\bibinfo {year} {1977})}\BibitemShut
  {NoStop}%
\bibitem [{\citenamefont {Scuseria}\ \emph {et~al.}(1987)\citenamefont
  {Scuseria}, \citenamefont {Scheiner}, \citenamefont {Lee}, \citenamefont
  {Rice},\ and\ \citenamefont {Schaefer~III}}]{scuseria1987closed}%
  \BibitemOpen
  \bibfield  {author} {\bibinfo {author} {\bibfnamefont {G.~E.}\ \bibnamefont
  {Scuseria}}, \bibinfo {author} {\bibfnamefont {A.~C.}\ \bibnamefont
  {Scheiner}}, \bibinfo {author} {\bibfnamefont {T.~J.}\ \bibnamefont {Lee}},
  \bibinfo {author} {\bibfnamefont {J.~E.}\ \bibnamefont {Rice}},\ and\
  \bibinfo {author} {\bibfnamefont {H.~F.}\ \bibnamefont {Schaefer~III}},\
  }\bibfield  {title} {\bibinfo {title} {The closed-shell coupled cluster
  single and double excitation (ccsd) model for the description of electron
  correlation. a comparison with configuration interaction (cisd) results},\
  }\href {https://aip.scitation.org/doi/10.1063/1.452039} {\bibfield  {journal}
  {\bibinfo  {journal} {J. Chem. Phys}\ }\textbf {\bibinfo {volume} {86}},\
  \bibinfo {pages} {2881} (\bibinfo {year} {1987})}\BibitemShut {NoStop}%
\bibitem [{\citenamefont {Szalay}\ and\ \citenamefont
  {Gauss}(1997)}]{szalay1997spin}%
  \BibitemOpen
  \bibfield  {author} {\bibinfo {author} {\bibfnamefont {P.~G.}\ \bibnamefont
  {Szalay}}\ and\ \bibinfo {author} {\bibfnamefont {J.}~\bibnamefont {Gauss}},\
  }\bibfield  {title} {\bibinfo {title} {Spin-restricted open-shell
  coupled-cluster theory},\ }\href
  {https://aip.scitation.org/doi/10.1063/1.475220} {\bibfield  {journal}
  {\bibinfo  {journal} {J. Chem. Phys}\ }\textbf {\bibinfo {volume} {107}},\
  \bibinfo {pages} {9028} (\bibinfo {year} {1997})}\BibitemShut {NoStop}%
\bibitem [{\citenamefont {Grimsley}\ \emph {et~al.}(2019)\citenamefont
  {Grimsley}, \citenamefont {Claudino}, \citenamefont {Economou}, \citenamefont
  {Barnes},\ and\ \citenamefont {Mayhall}}]{grimsley2019trotterized}%
  \BibitemOpen
  \bibfield  {author} {\bibinfo {author} {\bibfnamefont {H.~R.}\ \bibnamefont
  {Grimsley}}, \bibinfo {author} {\bibfnamefont {D.}~\bibnamefont {Claudino}},
  \bibinfo {author} {\bibfnamefont {S.~E.}\ \bibnamefont {Economou}}, \bibinfo
  {author} {\bibfnamefont {E.}~\bibnamefont {Barnes}},\ and\ \bibinfo {author}
  {\bibfnamefont {N.~J.}\ \bibnamefont {Mayhall}},\ }\bibfield  {title}
  {\bibinfo {title} {Is the trotterized {UCCSD} {A}nsatz chemically
  well-defined?},\ }\href {https://pubs.acs.org/doi/10.1021/acs.jctc.9b01083}
  {\bibfield  {journal} {\bibinfo  {journal} {J. Chem. Theory Comput}\ }\textbf
  {\bibinfo {volume} {16}},\ \bibinfo {pages} {1} (\bibinfo {year}
  {2019})}\BibitemShut {NoStop}%
\bibitem [{\citenamefont {Tsuchimochi}\ \emph {et~al.}(2020)\citenamefont
  {Tsuchimochi}, \citenamefont {Mori},\ and\ \citenamefont
  {Ten-no}}]{tsuchimochi2020spin}%
  \BibitemOpen
  \bibfield  {author} {\bibinfo {author} {\bibfnamefont {T.}~\bibnamefont
  {Tsuchimochi}}, \bibinfo {author} {\bibfnamefont {Y.}~\bibnamefont {Mori}},\
  and\ \bibinfo {author} {\bibfnamefont {S.~L.}\ \bibnamefont {Ten-no}},\
  }\bibfield  {title} {\bibinfo {title} {Spin-projection for quantum
  computation: A low-depth approach to strong correlation},\ }\href
  {https://journals.aps.org/prresearch/abstract/10.1103/PhysRevResearch.2.043142}
  {\bibfield  {journal} {\bibinfo  {journal} {Phys. Rev. Research}\ }\textbf
  {\bibinfo {volume} {2}},\ \bibinfo {pages} {043142} (\bibinfo {year}
  {2020})}\BibitemShut {NoStop}%
\bibitem [{\citenamefont {Izmaylov}\ \emph {et~al.}(2020)\citenamefont
  {Izmaylov}, \citenamefont {D{\'\i}az-Tinoco},\ and\ \citenamefont
  {Lang}}]{izmaylov2020order}%
  \BibitemOpen
  \bibfield  {author} {\bibinfo {author} {\bibfnamefont {A.~F.}\ \bibnamefont
  {Izmaylov}}, \bibinfo {author} {\bibfnamefont {M.}~\bibnamefont
  {D{\'\i}az-Tinoco}},\ and\ \bibinfo {author} {\bibfnamefont {R.~A.}\
  \bibnamefont {Lang}},\ }\bibfield  {title} {\bibinfo {title} {On the order
  problem in construction of unitary operators for the variational quantum
  eigensolver},\ }\href
  {https://pubs.rsc.org/en/content/articlelanding/2020/cp/d0cp01707h/unauth}
  {\bibfield  {journal} {\bibinfo  {journal} {Phys. Chem. Chem. Phys}\ }\textbf
  {\bibinfo {volume} {22}},\ \bibinfo {pages} {12980} (\bibinfo {year}
  {2020})}\BibitemShut {NoStop}%
\bibitem [{\citenamefont {LeBlanc}\ \emph {et~al.}(2015)\citenamefont {LeBlanc}
  \emph {et~al.}}]{LeBlanc_PRX_2015}%
  \BibitemOpen
  \bibfield  {author} {\bibinfo {author} {\bibfnamefont {J.~P.~F.}\
  \bibnamefont {LeBlanc}} \emph {et~al.},\ }\bibfield  {title} {\bibinfo
  {title} {Solutions of the two-dimensional {H}ubbard model: Benchmarks and
  results from a wide range of numerical algorithms},\ }\href
  {https://doi.org/10.1103/PhysRevX.5.041041} {\bibfield  {journal} {\bibinfo
  {journal} {Phys. Rev. X}\ }\textbf {\bibinfo {volume} {5}},\ \bibinfo {pages}
  {041041} (\bibinfo {year} {2015})}\BibitemShut {NoStop}%
\bibitem [{\citenamefont {Zheng}\ \emph {et~al.}(2017)\citenamefont {Zheng},
  \citenamefont {Chung}, \citenamefont {Corboz}, \citenamefont {Ehlers},
  \citenamefont {Qin}, \citenamefont {Noack}, \citenamefont {Shi},
  \citenamefont {White}, \citenamefont {Zhang},\ and\ \citenamefont
  {Chan}}]{Zheng_Science_2017}%
  \BibitemOpen
  \bibfield  {author} {\bibinfo {author} {\bibfnamefont {B.-X.}\ \bibnamefont
  {Zheng}}, \bibinfo {author} {\bibfnamefont {C.-M.}\ \bibnamefont {Chung}},
  \bibinfo {author} {\bibfnamefont {P.}~\bibnamefont {Corboz}}, \bibinfo
  {author} {\bibfnamefont {G.}~\bibnamefont {Ehlers}}, \bibinfo {author}
  {\bibfnamefont {M.-P.}\ \bibnamefont {Qin}}, \bibinfo {author} {\bibfnamefont
  {R.~M.}\ \bibnamefont {Noack}}, \bibinfo {author} {\bibfnamefont
  {H.}~\bibnamefont {Shi}}, \bibinfo {author} {\bibfnamefont {S.~R.}\
  \bibnamefont {White}}, \bibinfo {author} {\bibfnamefont {S.}~\bibnamefont
  {Zhang}},\ and\ \bibinfo {author} {\bibfnamefont {G.~K.-L.}\ \bibnamefont
  {Chan}},\ }\bibfield  {title} {\bibinfo {title} {Stripe order in the
  underdoped region of the two-dimensional {H}ubbard model},\ }\href
  {https://doi.org/10.1126/science.aam7127} {\bibfield  {journal} {\bibinfo
  {journal} {Science}\ }\textbf {\bibinfo {volume} {358}},\ \bibinfo {pages}
  {1155} (\bibinfo {year} {2017})}\BibitemShut {NoStop}%
\bibitem [{\citenamefont {Motta}\ \emph {et~al.}(2017)\citenamefont {Motta},
  \citenamefont {Ceperley}, \citenamefont {Chan}, \citenamefont {Gomez},
  \citenamefont {Gull}, \citenamefont {Guo}, \citenamefont {Jim\'enez-Hoyos},
  \citenamefont {Lan}, \citenamefont {Li}, \citenamefont {Ma}, \citenamefont
  {Millis}, \citenamefont {Prokof'ev}, \citenamefont {Ray}, \citenamefont
  {Scuseria}, \citenamefont {Sorella}, \citenamefont {Stoudenmire},
  \citenamefont {Sun}, \citenamefont {Tupitsyn}, \citenamefont {White},
  \citenamefont {Zgid},\ and\ \citenamefont {Zhang}}]{Motta_PRX_2017}%
  \BibitemOpen
  \bibfield  {author} {\bibinfo {author} {\bibfnamefont {M.}~\bibnamefont
  {Motta}}, \bibinfo {author} {\bibfnamefont {D.~M.}\ \bibnamefont {Ceperley}},
  \bibinfo {author} {\bibfnamefont {G.~K.-L.}\ \bibnamefont {Chan}}, \bibinfo
  {author} {\bibfnamefont {J.~A.}\ \bibnamefont {Gomez}}, \bibinfo {author}
  {\bibfnamefont {E.}~\bibnamefont {Gull}}, \bibinfo {author} {\bibfnamefont
  {S.}~\bibnamefont {Guo}}, \bibinfo {author} {\bibfnamefont {C.~A.}\
  \bibnamefont {Jim\'enez-Hoyos}}, \bibinfo {author} {\bibfnamefont {T.~N.}\
  \bibnamefont {Lan}}, \bibinfo {author} {\bibfnamefont {J.}~\bibnamefont
  {Li}}, \bibinfo {author} {\bibfnamefont {F.}~\bibnamefont {Ma}}, \bibinfo
  {author} {\bibfnamefont {A.~J.}\ \bibnamefont {Millis}}, \bibinfo {author}
  {\bibfnamefont {N.~V.}\ \bibnamefont {Prokof'ev}}, \bibinfo {author}
  {\bibfnamefont {U.}~\bibnamefont {Ray}}, \bibinfo {author} {\bibfnamefont
  {G.~E.}\ \bibnamefont {Scuseria}}, \bibinfo {author} {\bibfnamefont
  {S.}~\bibnamefont {Sorella}}, \bibinfo {author} {\bibfnamefont {E.~M.}\
  \bibnamefont {Stoudenmire}}, \bibinfo {author} {\bibfnamefont
  {Q.}~\bibnamefont {Sun}}, \bibinfo {author} {\bibfnamefont {I.~S.}\
  \bibnamefont {Tupitsyn}}, \bibinfo {author} {\bibfnamefont {S.~R.}\
  \bibnamefont {White}}, \bibinfo {author} {\bibfnamefont {D.}~\bibnamefont
  {Zgid}},\ and\ \bibinfo {author} {\bibfnamefont {S.}~\bibnamefont {Zhang}},\
  }\bibfield  {title} {\bibinfo {title} {Towards the solution of the
  many-electron problem in real materials: Equation of state of the {H}ydrogen
  chain with state-of-the-art many-body methods},\ }\href
  {https://doi.org/10.1103/PhysRevX.7.031059} {\bibfield  {journal} {\bibinfo
  {journal} {Phys. Rev. X}\ }\textbf {\bibinfo {volume} {7}},\ \bibinfo {pages}
  {031059} (\bibinfo {year} {2017})}\BibitemShut {NoStop}%
\bibitem [{\citenamefont {Williams}\ \emph {et~al.}(2020)\citenamefont
  {Williams}, \citenamefont {Yao}, \citenamefont {Li}, \citenamefont {Chen},
  \citenamefont {Shi}, \citenamefont {Motta}, \citenamefont {Niu},
  \citenamefont {Ray}, \citenamefont {Guo}, \citenamefont {Anderson} \emph
  {et~al.}}]{williams2020direct}%
  \BibitemOpen
  \bibfield  {author} {\bibinfo {author} {\bibfnamefont {K.~T.}\ \bibnamefont
  {Williams}}, \bibinfo {author} {\bibfnamefont {Y.}~\bibnamefont {Yao}},
  \bibinfo {author} {\bibfnamefont {J.}~\bibnamefont {Li}}, \bibinfo {author}
  {\bibfnamefont {L.}~\bibnamefont {Chen}}, \bibinfo {author} {\bibfnamefont
  {H.}~\bibnamefont {Shi}}, \bibinfo {author} {\bibfnamefont {M.}~\bibnamefont
  {Motta}}, \bibinfo {author} {\bibfnamefont {C.}~\bibnamefont {Niu}}, \bibinfo
  {author} {\bibfnamefont {U.}~\bibnamefont {Ray}}, \bibinfo {author}
  {\bibfnamefont {S.}~\bibnamefont {Guo}}, \bibinfo {author} {\bibfnamefont
  {R.~J.}\ \bibnamefont {Anderson}}, \emph {et~al.},\ }\bibfield  {title}
  {\bibinfo {title} {Direct comparison of many-body methods for realistic
  electronic {H}amiltonians},\ }\href
  {https://journals.aps.org/prx/abstract/10.1103/PhysRevX.10.011041} {\bibfield
   {journal} {\bibinfo  {journal} {Phys. Rev. X}\ }\textbf {\bibinfo {volume}
  {10}},\ \bibinfo {pages} {011041} (\bibinfo {year} {2020})}\BibitemShut
  {NoStop}%
\bibitem [{\citenamefont {Sun}\ \emph {et~al.}(2018)\citenamefont {Sun},
  \citenamefont {Berkelbach}, \citenamefont {Blunt}, \citenamefont {Booth},
  \citenamefont {Guo}, \citenamefont {Li}, \citenamefont {Liu}, \citenamefont
  {McClain}, \citenamefont {Sayfutyarova}, \citenamefont {Sharma} \emph
  {et~al.}}]{sun2018pyscf}%
  \BibitemOpen
  \bibfield  {author} {\bibinfo {author} {\bibfnamefont {Q.}~\bibnamefont
  {Sun}}, \bibinfo {author} {\bibfnamefont {T.~C.}\ \bibnamefont {Berkelbach}},
  \bibinfo {author} {\bibfnamefont {N.~S.}\ \bibnamefont {Blunt}}, \bibinfo
  {author} {\bibfnamefont {G.~H.}\ \bibnamefont {Booth}}, \bibinfo {author}
  {\bibfnamefont {S.}~\bibnamefont {Guo}}, \bibinfo {author} {\bibfnamefont
  {Z.}~\bibnamefont {Li}}, \bibinfo {author} {\bibfnamefont {J.}~\bibnamefont
  {Liu}}, \bibinfo {author} {\bibfnamefont {J.~D.}\ \bibnamefont {McClain}},
  \bibinfo {author} {\bibfnamefont {E.~R.}\ \bibnamefont {Sayfutyarova}},
  \bibinfo {author} {\bibfnamefont {S.}~\bibnamefont {Sharma}}, \emph
  {et~al.},\ }\bibfield  {title} {\bibinfo {title} {{P}y{SCF}: the python-based
  simulations of chemistry framework},\ }\href
  {https://onlinelibrary.wiley.com/doi/abs/10.1002/wcms.1340} {\bibfield
  {journal} {\bibinfo  {journal} {WIREs Comput. Mol. Sci}\ }\textbf {\bibinfo
  {volume} {8}},\ \bibinfo {pages} {e1340} (\bibinfo {year}
  {2018})}\BibitemShut {NoStop}%
\bibitem [{\citenamefont {Sun}\ \emph {et~al.}(2020)\citenamefont {Sun} \emph
  {et~al.}}]{sun2020recent}%
  \BibitemOpen
  \bibfield  {author} {\bibinfo {author} {\bibfnamefont {Q.}~\bibnamefont
  {Sun}} \emph {et~al.},\ }\bibfield  {title} {\bibinfo {title} {Recent
  developments in the pyscf program package},\ }\href
  {https://doi.org/10.1063/5.0006074} {\bibfield  {journal} {\bibinfo
  {journal} {J. Chem. Phys}\ }\textbf {\bibinfo {volume} {153}},\ \bibinfo
  {pages} {024109} (\bibinfo {year} {2020})}\BibitemShut {NoStop}%
\bibitem [{\citenamefont {Aleksandrowicz}\ \emph {et~al.}(2019)\citenamefont
  {Aleksandrowicz}, \citenamefont {Alexander}, \citenamefont {Barkoutsos},
  \citenamefont {Bello}, \citenamefont {Ben-Haim}, \citenamefont {Bucher},
  \citenamefont {Cabrera-Hern{\'a}ndez}, \citenamefont {Carballo-Franquis},
  \citenamefont {Chen}, \citenamefont {Chen} \emph
  {et~al.}}]{aleksandrowicz2019qiskit}%
  \BibitemOpen
  \bibfield  {author} {\bibinfo {author} {\bibfnamefont {G.}~\bibnamefont
  {Aleksandrowicz}}, \bibinfo {author} {\bibfnamefont {T.}~\bibnamefont
  {Alexander}}, \bibinfo {author} {\bibfnamefont {P.}~\bibnamefont
  {Barkoutsos}}, \bibinfo {author} {\bibfnamefont {L.}~\bibnamefont {Bello}},
  \bibinfo {author} {\bibfnamefont {Y.}~\bibnamefont {Ben-Haim}}, \bibinfo
  {author} {\bibfnamefont {D.}~\bibnamefont {Bucher}}, \bibinfo {author}
  {\bibfnamefont {F.}~\bibnamefont {Cabrera-Hern{\'a}ndez}}, \bibinfo {author}
  {\bibfnamefont {J.}~\bibnamefont {Carballo-Franquis}}, \bibinfo {author}
  {\bibfnamefont {A.}~\bibnamefont {Chen}}, \bibinfo {author} {\bibfnamefont
  {C.}~\bibnamefont {Chen}}, \emph {et~al.},\ }\bibfield  {title} {\bibinfo
  {title} {Qiskit: An open-source framework for quantum computing},\ }\href
  {https://zenodo.org/record/2562111#.XhA8qi2ZPyI} {\bibfield  {journal}
  {\bibinfo  {journal} {Zenodo}\ }\textbf {\bibinfo {volume} {16}} (\bibinfo
  {year} {2019})}\BibitemShut {NoStop}%
\bibitem [{\citenamefont {Zhu}\ \emph {et~al.}(1997)\citenamefont {Zhu},
  \citenamefont {Byrd}, \citenamefont {Lu},\ and\ \citenamefont
  {Nocedal}}]{zhu1997algorithm}%
  \BibitemOpen
  \bibfield  {author} {\bibinfo {author} {\bibfnamefont {C.}~\bibnamefont
  {Zhu}}, \bibinfo {author} {\bibfnamefont {R.~H.}\ \bibnamefont {Byrd}},
  \bibinfo {author} {\bibfnamefont {P.}~\bibnamefont {Lu}},\ and\ \bibinfo
  {author} {\bibfnamefont {J.}~\bibnamefont {Nocedal}},\ }\bibfield  {title}
  {\bibinfo {title} {Algorithm 778: {L-BFGS-B}: {F}ortran subroutines for
  large-scale bound-constrained optimization},\ }\href
  {https://doi.org/10.1145/279232.279236} {\bibfield  {journal} {\bibinfo
  {journal} {ACM Trans. Math. Softw.}\ }\textbf {\bibinfo {volume} {23}},\
  \bibinfo {pages} {550–560} (\bibinfo {year} {1997})}\BibitemShut {NoStop}%
\bibitem [{\citenamefont {Morales}\ and\ \citenamefont
  {Nocedal}(2011)}]{morales2011remark}%
  \BibitemOpen
  \bibfield  {author} {\bibinfo {author} {\bibfnamefont {J.~L.}\ \bibnamefont
  {Morales}}\ and\ \bibinfo {author} {\bibfnamefont {J.}~\bibnamefont
  {Nocedal}},\ }\bibfield  {title} {\bibinfo {title} {Remark on" algorithm 778:
  {L-BFGS-B}: {F}ortran subroutines for large-scale bound constrained
  optimization".},\ }\href {https://doi.org/10.1145/2049662.2049669} {\bibfield
   {journal} {\bibinfo  {journal} {ACM Trans. Math. Softw.}\ }\textbf {\bibinfo
  {volume} {38}},\ \bibinfo {pages} {7} (\bibinfo {year} {2011})}\BibitemShut
  {NoStop}%
\bibitem [{\citenamefont {Coulson}\ and\ \citenamefont
  {Fischer}(1949)}]{coulson1949xxxiv}%
  \BibitemOpen
  \bibfield  {author} {\bibinfo {author} {\bibfnamefont {C.~A.}\ \bibnamefont
  {Coulson}}\ and\ \bibinfo {author} {\bibfnamefont {I.}~\bibnamefont
  {Fischer}},\ }\bibfield  {title} {\bibinfo {title} {Xxxiv. notes on the
  molecular orbital treatment of the hydrogen molecule},\ }\href
  {https://www.tandfonline.com/doi/abs/10.1080/14786444908521726} {\bibfield
  {journal} {\bibinfo  {journal} {Philos. Mag}\ }\textbf {\bibinfo {volume}
  {40}},\ \bibinfo {pages} {386} (\bibinfo {year} {1949})}\BibitemShut
  {NoStop}%
\bibitem [{\citenamefont {Holmes}\ \emph {et~al.}(2022)\citenamefont {Holmes},
  \citenamefont {Sharma}, \citenamefont {Cerezo},\ and\ \citenamefont
  {Coles}}]{holmes2022connecting}%
  \BibitemOpen
  \bibfield  {author} {\bibinfo {author} {\bibfnamefont {Z.}~\bibnamefont
  {Holmes}}, \bibinfo {author} {\bibfnamefont {K.}~\bibnamefont {Sharma}},
  \bibinfo {author} {\bibfnamefont {M.}~\bibnamefont {Cerezo}},\ and\ \bibinfo
  {author} {\bibfnamefont {P.~J.}\ \bibnamefont {Coles}},\ }\bibfield  {title}
  {\bibinfo {title} {Connecting ansatz expressibility to gradient magnitudes
  and barren plateaus},\ }\href
  {https://journals.aps.org/prxquantum/abstract/10.1103/PRXQuantum.3.010313}
  {\bibfield  {journal} {\bibinfo  {journal} {PRX Quantum}\ }\textbf {\bibinfo
  {volume} {3}},\ \bibinfo {pages} {010313} (\bibinfo {year}
  {2022})}\BibitemShut {NoStop}%
\bibitem [{\citenamefont {D'Cunha}\ \emph {et~al.}(2022)\citenamefont
  {D'Cunha}, \citenamefont {Crawford}, \citenamefont {Motta},\ and\
  \citenamefont {Rice}}]{github}%
  \BibitemOpen
  \bibfield  {author} {\bibinfo {author} {\bibfnamefont {R.}~\bibnamefont
  {D'Cunha}}, \bibinfo {author} {\bibfnamefont {T.~D.}\ \bibnamefont
  {Crawford}}, \bibinfo {author} {\bibfnamefont {M.}~\bibnamefont {Motta}},\
  and\ \bibinfo {author} {\bibfnamefont {J.~E.}\ \bibnamefont {Rice}},\ }\href
  {https://github.com/mariomotta/quantum_database} {\bibinfo {title} {Github
  repository}} (\bibinfo {year} {2022})\BibitemShut {NoStop}%
\bibitem [{\citenamefont {Parrish}\ \emph {et~al.}(2019)\citenamefont
  {Parrish}, \citenamefont {Hohenstein}, \citenamefont {McMahon},\ and\
  \citenamefont {Mart{\'\i}nez}}]{parrish2019quantum}%
  \BibitemOpen
  \bibfield  {author} {\bibinfo {author} {\bibfnamefont {R.~M.}\ \bibnamefont
  {Parrish}}, \bibinfo {author} {\bibfnamefont {E.~G.}\ \bibnamefont
  {Hohenstein}}, \bibinfo {author} {\bibfnamefont {P.~L.}\ \bibnamefont
  {McMahon}},\ and\ \bibinfo {author} {\bibfnamefont {T.~J.}\ \bibnamefont
  {Mart{\'\i}nez}},\ }\bibfield  {title} {\bibinfo {title} {Quantum computation
  of electronic transitions using a variational quantum eigensolver},\ }\href
  {https://journals.aps.org/prl/abstract/10.1103/PhysRevLett.122.230401}
  {\bibfield  {journal} {\bibinfo  {journal} {Phys. Rev. Lett}\ }\textbf
  {\bibinfo {volume} {122}},\ \bibinfo {pages} {230401} (\bibinfo {year}
  {2019})}\BibitemShut {NoStop}%
\bibitem [{\citenamefont {Anselmetti}\ \emph {et~al.}(2021)\citenamefont
  {Anselmetti}, \citenamefont {Wierichs}, \citenamefont {Gogolin},\ and\
  \citenamefont {Parrish}}]{anselmetti2021local}%
  \BibitemOpen
  \bibfield  {author} {\bibinfo {author} {\bibfnamefont {G.-L.~R.}\
  \bibnamefont {Anselmetti}}, \bibinfo {author} {\bibfnamefont
  {D.}~\bibnamefont {Wierichs}}, \bibinfo {author} {\bibfnamefont
  {C.}~\bibnamefont {Gogolin}},\ and\ \bibinfo {author} {\bibfnamefont {R.~M.}\
  \bibnamefont {Parrish}},\ }\bibfield  {title} {\bibinfo {title} {Local,
  expressive, quantum-number-preserving vqe ans{\"a}tze for fermionic
  systems},\ }\href
  {https://iopscience.iop.org/article/10.1088/1367-2630/ac2cb3/pdf} {\bibfield
  {journal} {\bibinfo  {journal} {New J. Phys}\ }\textbf {\bibinfo {volume}
  {23}},\ \bibinfo {pages} {113010} (\bibinfo {year} {2021})}\BibitemShut
  {NoStop}%
\bibitem [{\citenamefont {McClean}\ \emph {et~al.}(2017)\citenamefont
  {McClean}, \citenamefont {Kimchi-Schwartz}, \citenamefont {Carter},\ and\
  \citenamefont {De~Jong}}]{mcclean2017hybrid}%
  \BibitemOpen
  \bibfield  {author} {\bibinfo {author} {\bibfnamefont {J.~R.}\ \bibnamefont
  {McClean}}, \bibinfo {author} {\bibfnamefont {M.~E.}\ \bibnamefont
  {Kimchi-Schwartz}}, \bibinfo {author} {\bibfnamefont {J.}~\bibnamefont
  {Carter}},\ and\ \bibinfo {author} {\bibfnamefont {W.~A.}\ \bibnamefont
  {De~Jong}},\ }\bibfield  {title} {\bibinfo {title} {Hybrid quantum-classical
  hierarchy for mitigation of decoherence and determination of excited
  states},\ }\href
  {https://journals.aps.org/pra/abstract/10.1103/PhysRevA.95.042308} {\bibfield
   {journal} {\bibinfo  {journal} {Phys. Rev. A}\ }\textbf {\bibinfo {volume}
  {95}},\ \bibinfo {pages} {042308} (\bibinfo {year} {2017})}\BibitemShut
  {NoStop}%
\bibitem [{\citenamefont {Huggins}\ \emph {et~al.}(2020)\citenamefont
  {Huggins}, \citenamefont {Lee}, \citenamefont {Baek}, \citenamefont
  {O’Gorman},\ and\ \citenamefont {Whaley}}]{huggins2020non}%
  \BibitemOpen
  \bibfield  {author} {\bibinfo {author} {\bibfnamefont {W.~J.}\ \bibnamefont
  {Huggins}}, \bibinfo {author} {\bibfnamefont {J.}~\bibnamefont {Lee}},
  \bibinfo {author} {\bibfnamefont {U.}~\bibnamefont {Baek}}, \bibinfo {author}
  {\bibfnamefont {B.}~\bibnamefont {O’Gorman}},\ and\ \bibinfo {author}
  {\bibfnamefont {K.~B.}\ \bibnamefont {Whaley}},\ }\bibfield  {title}
  {\bibinfo {title} {A non-orthogonal variational quantum eigensolver},\ }\href
  {https://iopscience.iop.org/article/10.1088/1367-2630/ab867b} {\bibfield
  {journal} {\bibinfo  {journal} {New J. Phys}\ }\textbf {\bibinfo {volume}
  {22}},\ \bibinfo {pages} {073009} (\bibinfo {year} {2020})}\BibitemShut
  {NoStop}%
\bibitem [{\citenamefont {Stair}\ \emph {et~al.}(2020)\citenamefont {Stair},
  \citenamefont {Huang},\ and\ \citenamefont
  {Evangelista}}]{stair2020multireference}%
  \BibitemOpen
  \bibfield  {author} {\bibinfo {author} {\bibfnamefont {N.~H.}\ \bibnamefont
  {Stair}}, \bibinfo {author} {\bibfnamefont {R.}~\bibnamefont {Huang}},\ and\
  \bibinfo {author} {\bibfnamefont {F.~A.}\ \bibnamefont {Evangelista}},\
  }\bibfield  {title} {\bibinfo {title} {A multireference quantum {K}rylov
  algorithm for strongly correlated electrons},\ }\href
  {https://pubs.acs.org/doi/10.1021/acs.jctc.9b01125} {\bibfield  {journal}
  {\bibinfo  {journal} {J. Chem. Theory Comput}\ }\textbf {\bibinfo {volume}
  {16}},\ \bibinfo {pages} {2236} (\bibinfo {year} {2020})}\BibitemShut
  {NoStop}%
\bibitem [{\citenamefont {Cohn}\ \emph {et~al.}(2021)\citenamefont {Cohn},
  \citenamefont {Motta},\ and\ \citenamefont {Parrish}}]{cohn2021quantum}%
  \BibitemOpen
  \bibfield  {author} {\bibinfo {author} {\bibfnamefont {J.}~\bibnamefont
  {Cohn}}, \bibinfo {author} {\bibfnamefont {M.}~\bibnamefont {Motta}},\ and\
  \bibinfo {author} {\bibfnamefont {R.~M.}\ \bibnamefont {Parrish}},\
  }\bibfield  {title} {\bibinfo {title} {Quantum filter diagonalization with
  compressed double-factorized hamiltonians},\ }\href
  {https://doi.org/10.1103/PRXQuantum.2.040352} {\bibfield  {journal} {\bibinfo
   {journal} {PRX Quantum}\ }\textbf {\bibinfo {volume} {2}},\ \bibinfo {pages}
  {040352} (\bibinfo {year} {2021})}\BibitemShut {NoStop}%
\bibitem [{\citenamefont {Ditchfield}\ \emph {et~al.}(1971)\citenamefont
  {Ditchfield}, \citenamefont {Hehre},\ and\ \citenamefont
  {Pople}}]{ditchfield1971self}%
  \BibitemOpen
  \bibfield  {author} {\bibinfo {author} {\bibfnamefont {R.}~\bibnamefont
  {Ditchfield}}, \bibinfo {author} {\bibfnamefont {W.~J.}\ \bibnamefont
  {Hehre}},\ and\ \bibinfo {author} {\bibfnamefont {J.~A.}\ \bibnamefont
  {Pople}},\ }\bibfield  {title} {\bibinfo {title} {{Self-consistent
  molecular-orbital methods. {IX}. An extended {G}aussian-type basis for
  molecular-orbital studies of organic molecules}},\ }\href
  {https://aip.scitation.org/doi/10.1063/1.1674902} {\bibfield  {journal}
  {\bibinfo  {journal} {J. Chem. Phys}\ }\textbf {\bibinfo {volume} {54}},\
  \bibinfo {pages} {724} (\bibinfo {year} {1971})}\BibitemShut {NoStop}%
\bibitem [{\citenamefont {Dunning~Jr}(1989)}]{dunning1989gaussian}%
  \BibitemOpen
  \bibfield  {author} {\bibinfo {author} {\bibfnamefont {T.~H.}\ \bibnamefont
  {Dunning~Jr}},\ }\bibfield  {title} {\bibinfo {title} {Gaussian basis sets
  for use in correlated molecular calculations. i. the atoms boron through neon
  and hydrogen},\ }\href {https://aip.scitation.org/doi/10.1063/1.456153}
  {\bibfield  {journal} {\bibinfo  {journal} {J. Chem. Phys}\ }\textbf
  {\bibinfo {volume} {90}},\ \bibinfo {pages} {1007} (\bibinfo {year}
  {1989})}\BibitemShut {NoStop}%
\end{thebibliography}%

%\begin{tocentry}
%\centering
%\includegraphics[height=3.5cm]{graphical_toc_jpca.pdf}
%\end{tocentry}

\end{document}